\documentclass[preprint,3p,11pt]{elsarticle}

\usepackage{graphicx,amssymb}
\usepackage{epsf,psfig}

\journal{Astronomical \& Astrophysical Transactions}

\begin{document}

\begin{frontmatter}

\title{Using new dynamical indicators to distinguish between order and chaos \\
in a galactic potential producing exact periodic orbits \\
and chaotic components}

\author{Euaggelos E. Zotos\fnref{}}

\address{Department of Physics, \\
Section of  Astrophysics, Astronomy and Mechanics, \\
Aristotle University of Thessaloniki \\
GR-541 24, Thessaloniki, Greece}

\fntext[]{Corresponding author: \\
E-mail: evzotos@physics.auth.gr}

\begin{abstract}
The theory of the inverse problem is used in order to find a two dimensional galactic potential generating a mono-parametric family of elliptic periodic orbits. The potential is made up of a two-dimensional harmonic oscillator with perturbing terms of third and fourth degree and can be considered to describe local motion in the central parts of a barred galaxy. Our numerical calculations indicate that the potential produces also other families of orbits and there are cases, where distinct chaotic components are also observed. The numerical experiments suggest that there are regular as well as chaotic orbits supporting the barred structure. We use a variety of dynamical indicators, in order to determine the character of motion. Of significant interest, is the $S(g)$ dynamical spectrum, which is used in order to identify the islandic motion of resonant orbits and also proves to be a very useful and fast indicator, in order to distinguish between regular and chaotic motion. Comparison with other methods for detecting chaos is also discussed. The present results are compared with outcomes from earlier work.
\end{abstract}


\end{frontmatter}

\section{Introduction}

It is well known, that potentials based on a two-dimensional perturbed harmonic oscillator have been extensively used for about five decades in order to describe motion in galaxies (see H\'{e}non \& Heiles, 1964; Caranicolas, 1984; Caranicolas \& Innanen, 1992; Lara, 1996; Elipe \& Deprit, 1999; Arribas et al., 2006; Karanis \& Vozikis, 2008).

The general form of such a potential is
\begin{equation}
V = \frac{1}{2}\left(\omega_1^2 x^2 + \omega_2^2 y^2\right) + V_1,
\end{equation}
where $\omega_1$ and $\omega_2$ are the unperturbed frequencies along the $x$ and $y$ axis respectively, while $V_1$ is a polynomial containing the perturbing terms. A common way to understand the dynamical properties of the potential (1) is to find the various types of orbits associated with the potential. Of particular interest are the periodic orbits and their stability. Quasi periodic orbits are trapped in the vicinity of stable periodic orbits having the same topological features as the parent periodic orbits. Furthermore, unstable periodic orbits give rise to chaotic motion.

There are two basic reasons supporting our choice of the potential (1). The first reason is, that such a potential describes in a satisfactory way the properties of local motion in the central parts of a galaxy. The second and most important reason, is that this potential is easy to handle, not only numerically but also analytically in order to obtain interesting outcomes and compare between the numerical and the corresponding analytical results.

The procedure described above, that is to find the orbits in a given potential, is the well known ``direct problem" of dynamics. On the other hand, there is also the ``inverse problem" of dynamics, which seeks for potentials associated with a given family of orbits. In particular, the planar inverse problem of dynamics looks for potentials $V(x,y)$, which can produce a preassigned mono-parametric family of orbits. Families of orbits in planar anisotropic potentials have been found by Anisiu and Bozis (2004). Moreover, in the 3D inverse problem of dynamics we search for potentials $V(x,y,z)$, which can produce a two-parametric family of orbits (see Anisiu, 2005; Bozis \& Kotoulas, 2005; Anisiu \& Kotoulas, 2006; Kotoulas \& Bozis, 2006). The reader can also find interesting information concerning the Inverse Problem of Dynamics (IPD) in a series of papers (see Puel, 1984; 1992; Bozis \& Nakhla, 1986; Shorokhov, 1988; Bozis, 1995; Anisiu \& Pal, 1999).

In an earlier paper (Caranicolas, 2000), we have found polynomial potentials producing mono-parametric families of periodic orbits. In those potentials the part $V_1$ included only three terms of fourth degree or two terms of third and one term of fourth degree. Furthermore, all mono-parametric families of orbits were based only in one constant. In the present article, we shall look for a galactic potential with perturbing terms of third and fourth degree, which produces a mono-parametric family of ellipses based on two constants. The article is organized as follows: In Section 2 we use the theory of the inverse problem in order to construct the potential. In Section 3 the structure of the potential is examined through some analytical calculations and the astronomical perspective of the potential is outlined. In Section 4 the numerical exploration of all the families of orbits is presented and the conditions of the coexistence of both accurate periodic orbits and chaos are investigated, using a variety of dynamical indicators. In Section 5 we present a discussion and the conclusions of our research.

\section{The potential associated with the family of accurate periodic orbits}

We look for a potential of the general form
\begin{equation}
V(x,y) = \frac{1}{2}\left(x^2 + b y^2\right)
-\left[\alpha_1 y^3 + \alpha_2 x^2 y + \alpha_3 y^4 + \alpha_4 x^2 y^2 + \alpha_5 x^4\right],
\end{equation}
creating the mono-parametric family of ellipses
\begin{equation}
f(x,y) = A x^2 + B y + y^2 = c,
\end{equation}
where $A > 1$ and  $B$ are given constants, $c > 0$ is the parameter characterizing the family, while $b$, $\alpha_1$, $\alpha_2$, $\alpha_3$, $\alpha_4$ and $\alpha_5$ are constants to be determined. All potentials creating, among others, the family of orbits (3) are given by the second order partial differential equation (Bozis, 1984)
\begin{equation}
-V_{xx} + k V_{xy} + V_{yy} = \lambda V_x + \mu V_y,
\end{equation}
where
\begin{equation}
k = \frac{1 - \gamma^2}{\gamma}, \ \ \lambda = \frac{\Gamma_y - \gamma
  \Gamma_x}{\gamma \Gamma}, \ \ \mu = k \gamma + \frac{3\Gamma}{\gamma},
\end{equation}
with
\begin{equation}
\gamma = \frac{f_y}{f_x}, \ \ \Gamma = \gamma \gamma_x - \gamma_y.
\end{equation}
Each orbit, being a member of the family (3), has a total energy $E = E(c)$ which is given by Szebehely's equation
\begin{equation}
E = V(x,y) - \frac{1 + \gamma^2}{2\Gamma}\left(V_x + \gamma V_y\right).
\end{equation}
The inequality
\begin{equation}
\frac{1}{\Gamma}\left(V_x + \gamma V_y\right) \leq 0,
\end{equation}
defines the region on the $(x,y)$ plane (Bozis \& Ichtiaroglou, 1994), where motion is allowed for the members of the family (3). If we insert expressions (2) and (3) in equation (4) and demand that all coefficients of equal powers in $x$ and $y$ be zero we find
\begin{eqnarray}
&b& = \frac{2(1 + 2A)}{A - 1}, \ \ \ \alpha_1 = \frac{2(1 + 2A)}{(1 - A)B}, \nonumber \\
&\alpha_2& = \frac{6A}{(1 - A)B}, \ \ \ \alpha_3 = \frac{1 + 2A}{(1 - A)B^2}, \nonumber \\
&\alpha_4& = \frac{6A}{(1 - A)B^2}, \ \ \ \alpha_5 = \frac{(2 + A)A}{(1 - A)B^2}.
\end{eqnarray}
The total energy of the test particle of a unit mass moving on the each ellipse (3) is
\begin{equation}
E_{el} = \frac{\left(1 + 2A\right)\left(2A c + 4c + B^2\right)c}{2A\left(A - 1\right)B^2},
\end{equation}
while the region on the $(x,y)$ plane where motion is allowed is defined by
\begin{equation}
\left(B + 2x\right)^2[\left(1 + 2A\right)x^2 + Bx\left(1 + 2A\right)+ 
A\left(2 + A\right)y^2] \geq A(A - 1)B^2 \ .
\end{equation}

\section{The structure of the potential}

In this Section, we will examine the structure of the potential (2) with the values of its parameters given by Eq. (9). Hereafter, we shall call this potential $V_G(x,y)$. Since the potential creates the family of the ellipses (3), we shall seek for the conditions, where these ellipses and all other families of regular orbits created by this potential, can create and support a barred structure in the central region of the galaxy. Furthermore, we shall look if there are any chaotic orbits and their relation with the above mentioned barred structure.
\begin{figure}[!tH]
\centering
\resizebox{0.60\hsize}{!}{\rotatebox{0}{\includegraphics*{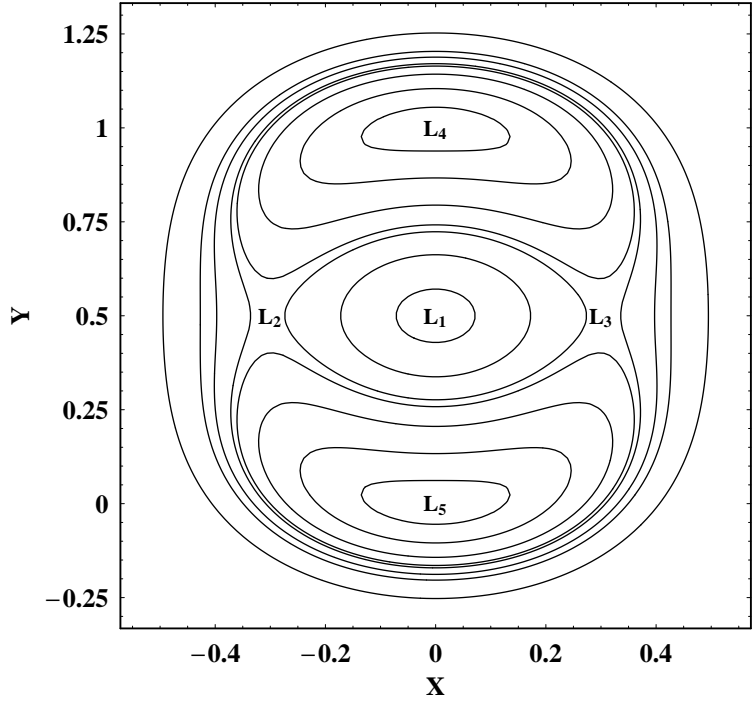}}}
\caption{Contours of constant potential $V_G(x,y) = const.$, when $A=4$ and $B=-1$. Here $L_1$, $L_2$, $L_3$, $L_4$ and $L_5$ are the five equilibrium points.}
\end{figure}
\begin{figure}[!tH]
\centering
\resizebox{0.60\hsize}{!}{\rotatebox{0}{\includegraphics*{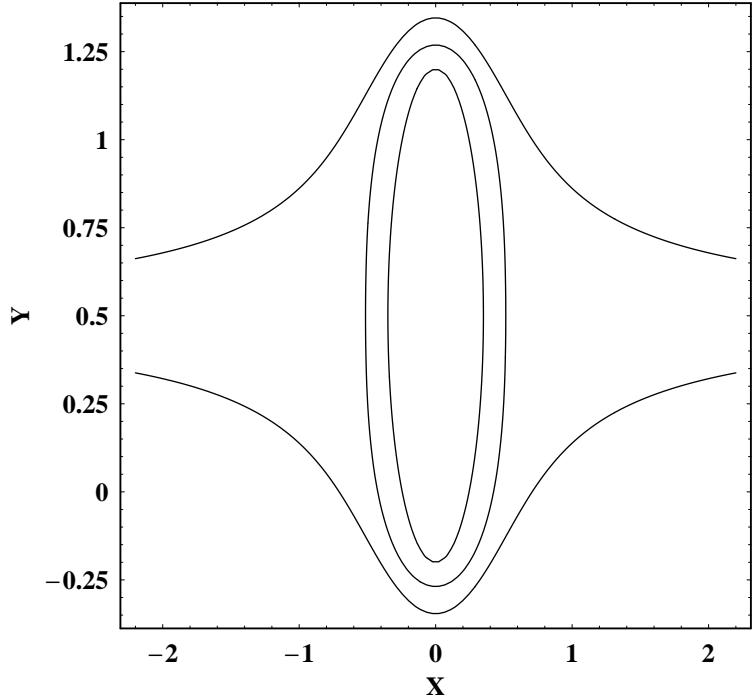}}}
\caption{From inside outwards: the elliptic accurate periodic orbit, the zero velocity curve and the family boundary curve. The values of the parameters are $A=4$, $B=-1$ and $c=0.24$}
\end{figure}

Let us first examine the structure of the potential $V_G$. The five equilibrium points $(L_1, L_2, L_3, L_4, L_5)$ of the potential, which are the solutions of the system of equations
\begin{equation}
\frac{\partial V_G}{\partial x} = 0, \ \ \frac{\partial V_G}{\partial y} = 0,
\end{equation}
are at
\begin{eqnarray}
(i). &\mathbf{L_1}&: \left(x = 0, \ \ y = -B/2\right), \nonumber \\
(ii). &\mathbf{L_2}&: \left(x = B \frac{\sqrt{1 + 2A}}{\sqrt{8A + 4A^2}}, \ \ y = -B/2\right), \nonumber \\
(iii). &\mathbf{L_3}&: \left(x = -B \frac{\sqrt{1 + 2A}}{\sqrt{8A + 4A^2}}, \ \ y = -B/2\right), \nonumber \\
(iv). &\mathbf{L_4}&: \left(x = 0, \ \ y = -B\right), \nonumber \\
(v). &\mathbf{L_5}&: \left(x = 0, \ \ y = 0\right). \ \ \
\end{eqnarray}
The value of $V_G$ at (i) is
\begin{equation}
V_{G1} = \frac{\left(1 + 2A\right)B^2}{16\left(A - 1\right)}.
\end{equation}
At (ii) and (iii) we have
\begin{equation}
V_{G2} = V_{G3} = \frac{\left(1 + 3A + 2A^2\right)B^2}{16A\left(2 + A\right)},
\end{equation}
while at (iv) and (v) we have
\begin{equation}
V_{G4} = V_{G5} = 0.
\end{equation}
It is interesting to see how the topology of the equipotential curves
\begin{equation}
V_G(x,y) = const.,
\end{equation}
changes as the value of the constant in (17) changes. This is very important from the astronomical point of view because the motion takes place inside the area defined by the equipotential curve, which is also known as the Zero Velocity Curve (ZVC).

An elementary analysis shows that for values of the energy $E$ in the range
\begin{equation}
0 < E \leq V_{G2,3},
\end{equation}
the ZVC is composed of two different areas of motion. On the other hand, for values of the energy $E$ in the range
\begin{equation}
V_{G4,5} < E < V_{G1},
\end{equation}
the above two regions merge and the ZVC is composed of one unified region, which looks like a torus. Finally, when
\begin{equation}
E \geq V_{G1} = \frac{\left(1 + 2A\right)B^2}{16\left(A - 1\right)},
\end{equation}
we have a ZVC which looks like a bar along the $y$-axis (see Fig. 2). Figure 1 shows the contours (17) when $A=4$ and $B=-1$. In this case we have $V_{G2,3} = 0.117188$ and $V_{G1} = 0.1875$. The contours are (0.01, 0.04, 0.08, 0.11, 0.12, 0.15, 0.18, 0.30), while $L_1$, $L_2$, $L_3$, $L_4$ and $L_5$ are the five equilibrium points.

As we are interested in a potential with barred structure, the energy must fulfill relation (20). On the other hand, the potential creates, among others, the family of the ellipses (3). For a triad of values of $A$, $B$ and $c$ we have an ellipse,
which must lie inside the corresponding ZVC. As the energy of the accurate elliptic periodic orbit is given by relation (10) we must always have
\begin{equation}
E_{el} \geq V_{G1}.
\end{equation}
Inserting the values of $E_{el}$ and $V_{G1}$ from relations (10) and (20) in equation (21) we obtain
\begin{equation}
8\left(2A c + 4c + B^2\right)c - A B^4 \geq 0.
\end{equation}
Solving equation (22) for $c$ we find
\begin{equation}
c \geq \frac{A B^2}{4(2 + A)}, \ \ c \leq - \frac{B^2}{2}.
\end{equation}

As we considered positive values of $c$, only the first solution (23) is acceptable. This relation connects $A$, $B$ and $c$ so that the ellipse (3) lies inside the corresponding ZVC. Figure 2 shows from inside outwards: the elliptic accurate periodic orbit, the zero velocity curve and the family boundary curve given by equation (11). The values of the constants are $A=4$, $B=-1$, while $c = 0.24$.

\section{Numerical experiments and the nature of the orbits}

In this Section, we shall use the numerical integration of the equations of motion corresponding to the Hamiltonian of the potential (2), which is
\begin{equation}
H_G = \frac{1}{2}\left(p_x^2 + p_y^2\right) + V_G,
\end{equation}
where $p_x$ and $p_y$ are the momenta per unit mass conjugate to $x$ and $y$ respectively. Our aim is to see if there are other families of orbits, in addition to the family of the ellipses (3). In order to see this we used the $(y, p_y)$, $(x=0, p_x > 0)$ Poincar\'{e} phase plane. In the following, we shall investigate the regular or chaotic character of orbits, using three different criteria. The first is the classical method of Lyapunov Characteristic Exponent - LCE (see Lichtenberg \& Liebermann, 1992), the second is the $S(g)$ dynamical spectrum and the third is the $P(f)$ spectral method used by Karanis and Vozikis (2007).

Dynamical spectra of orbits have been frequently used in galactic dynamics, as fast indicators of regular or chaotic motion (Caranicolas and Zotos, 2010; Zotos, 2011a; 2011b; 2011c; 2012). In an earlier work (Zotos, 2011a), we introduced and used a new dynamical parameter the $S(g)$ spectrum, in order to study the islandic motion of resonant orbits and also the evolution of sticky orbits. The $S(g)$ spectrum was applied in a axially symmetric galactic model in the meridian $(r, z)$ plane. However, it is easy enough to modify its definition in order to make
the $S(g)$ spectrum applicable in the Cartesian coordinates $(x, y)$. Thus, we define the dynamical parameter $g$ as
\begin{equation}
g_i=\frac{y_i + p_{yi} - y_i p_{yi}}{p_{xi}},
\end{equation}
where $(y_i, p_{xi}, p_{yi})$ are the successive values of the $(y, p_x, p_y)$ elements of a 2D orbit on the Poincar\'{e} $(y, p_y)$, $(x=0, p_x > 0)$ phase plane. The dynamical spectrum of the parameter $g$ is its distribution function
\begin{equation}
S(g)=\frac{\Delta N(g)}{N \Delta g},
\end{equation}
where $\Delta N(g)$ are the number of the parameters $g$ in the interval $(g, g + \Delta g)$, after $N$ iterations. By definition, the $g$ parameter is based on a combination of the coordinates and the velocities of a 2D orbit and therefore it is a combined dynamical spectrum.
\begin{figure}[!tH]
\centering
\resizebox{0.60\hsize}{!}{\rotatebox{270}{\includegraphics*{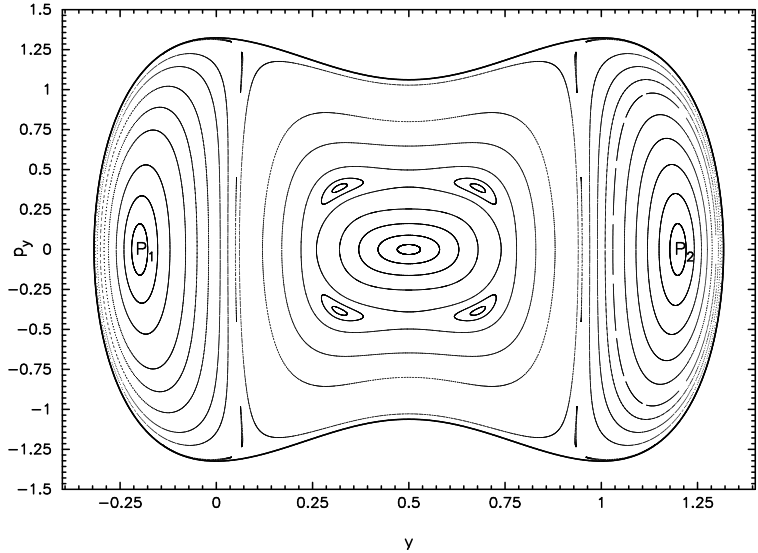}}}
\caption{The $(y, p_y)$ phase plane when $A=2$, $B=-1$ and $c=0.24$. The value of energy is $E_{el} = 0.8760$. $P1$ and $P2$ represent the two identical ellipses (3) traversed in opposite directions.}
\end{figure}

The $P(f)$ spectral method, uses the Fast Fourier Transform (FFT) of a series of time intervals, each one representing the time that elapsed between two successive points on the Poincar\'{e} $(y, p_y)$, $(x = 0, p_x > 0)$ surface of section, of the 2D dynamical system.

Figure 3 shows the $(y, p_y)$ phase plane when $A=2$, $B=-1$ and $c=0.24$. The value of the energy, which is found using equation (10) is $E_{el} = 0.8760$. The two points $P1$ and $P2$ represent the two identical ellipses (3) traversed in opposite directions. As we see, these ellipses are stable periodic orbits and also orbits starting near these ellipses, form quasi periodic tube orbits. Such a tube orbit is shown in Figure 4a. The initial conditions are: $y_0=1.05, x_0 = p_{y0}=0$, while the value of $p_{x0}$ is found using the energy integral in all cases. Figure 4b shows the corresponding LCE of the orbit, which was computed for a period of $10^5$ time units and it vanishes as expected. In Figure 4c, we can see a well defined $U$-type spectrum indicating regular motion, while in Figure 4d, the $P(f)$ indicator produces a small number of peaks. On the other hand, there are also families of box orbits forming invariant curves surrounding the central invariant point at $y = -B/2, p_y = 0$. This point represents a periodic orbit, which is a straight line parallel to the $x$ axis. Figure 5a shows a box orbit, while in Figures 5b, 5c and 5d we can see the corresponding LCE, the $S(g)$ spectrum and the $P(f)$ indicator respectively. The Initial conditions are: $y_0=0.85, x_0 = p_{y0}=0$. The motion is regular, while both families of the above mentioned orbits support the barred structure. In addition to the above families of orbits there are also orbits producing the set of small islands in the central region of Fig. 3.
\begin{figure*}[!tH]
\centering
\resizebox{0.70\textwidth}{!}{\rotatebox{0}{\includegraphics*{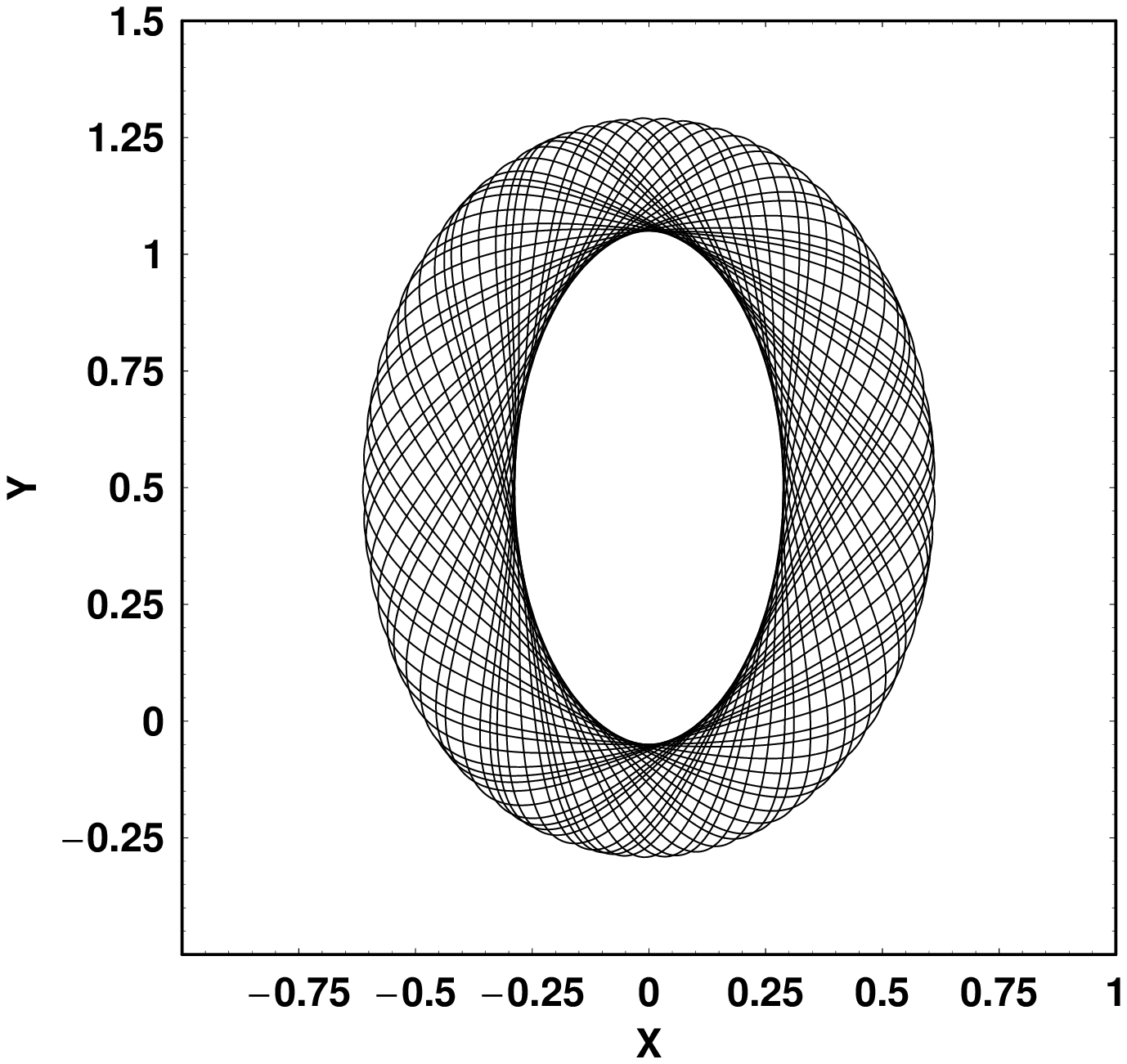}}\hspace{1cm}
                              \rotatebox{0}{\includegraphics*{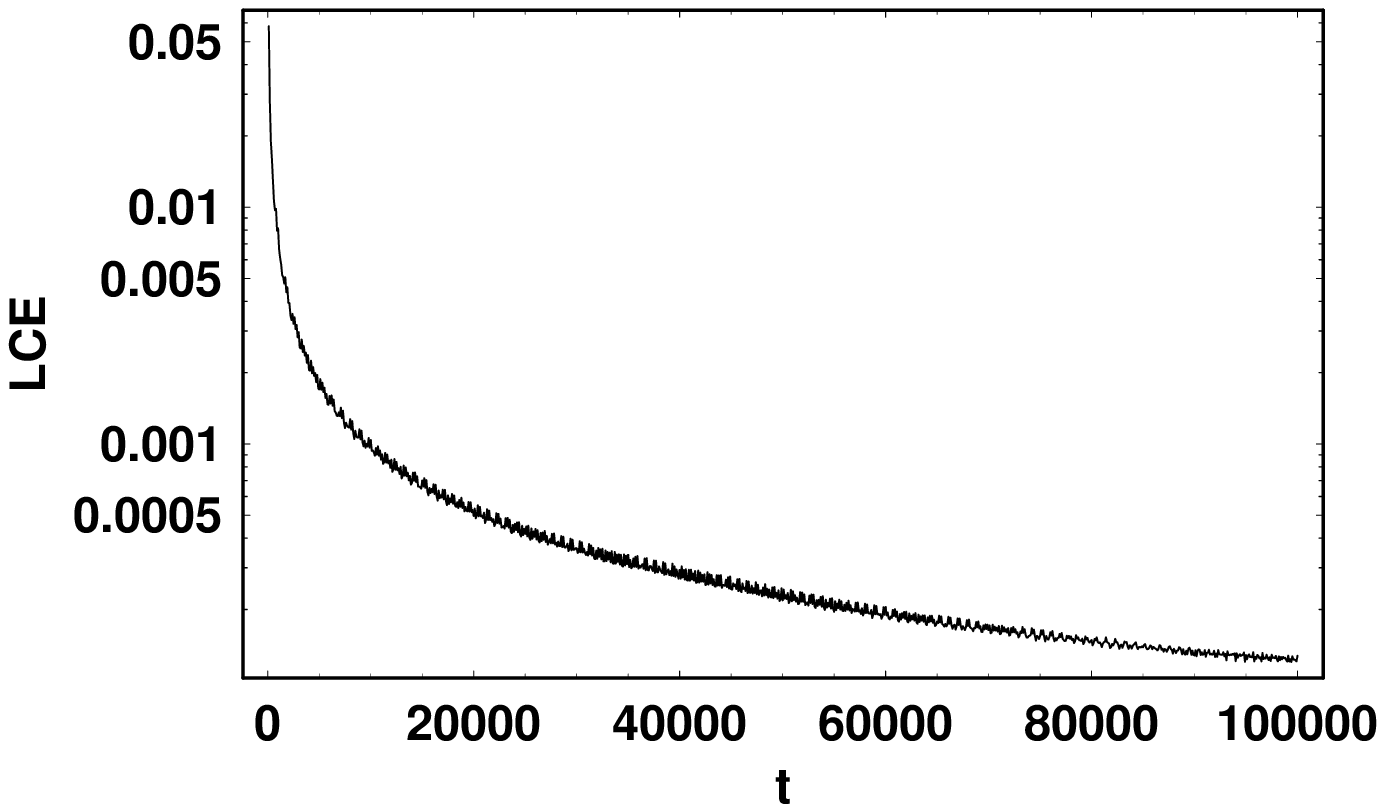}}}
\resizebox{0.70\textwidth}{!}{\rotatebox{0}{\includegraphics*{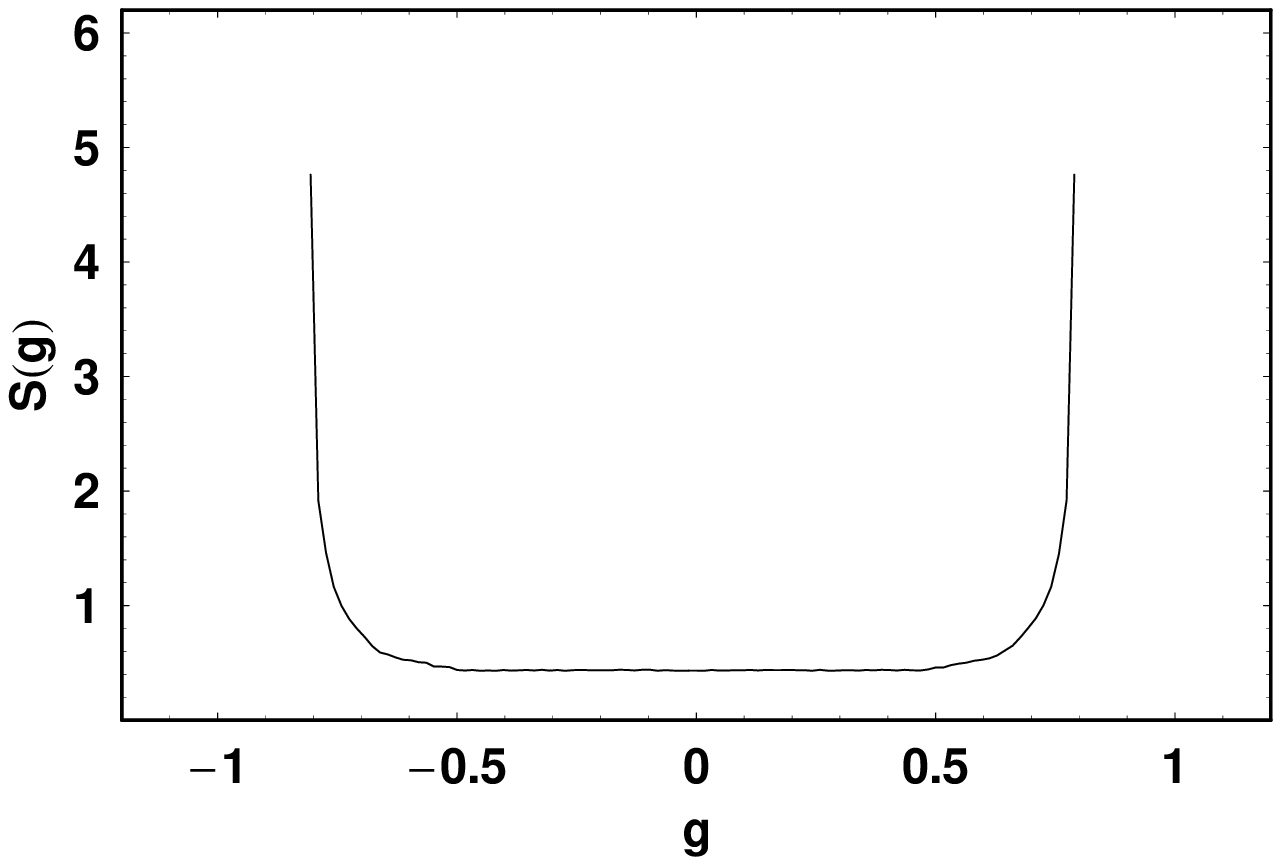}}\hspace{1cm}
                              \rotatebox{0}{\includegraphics*{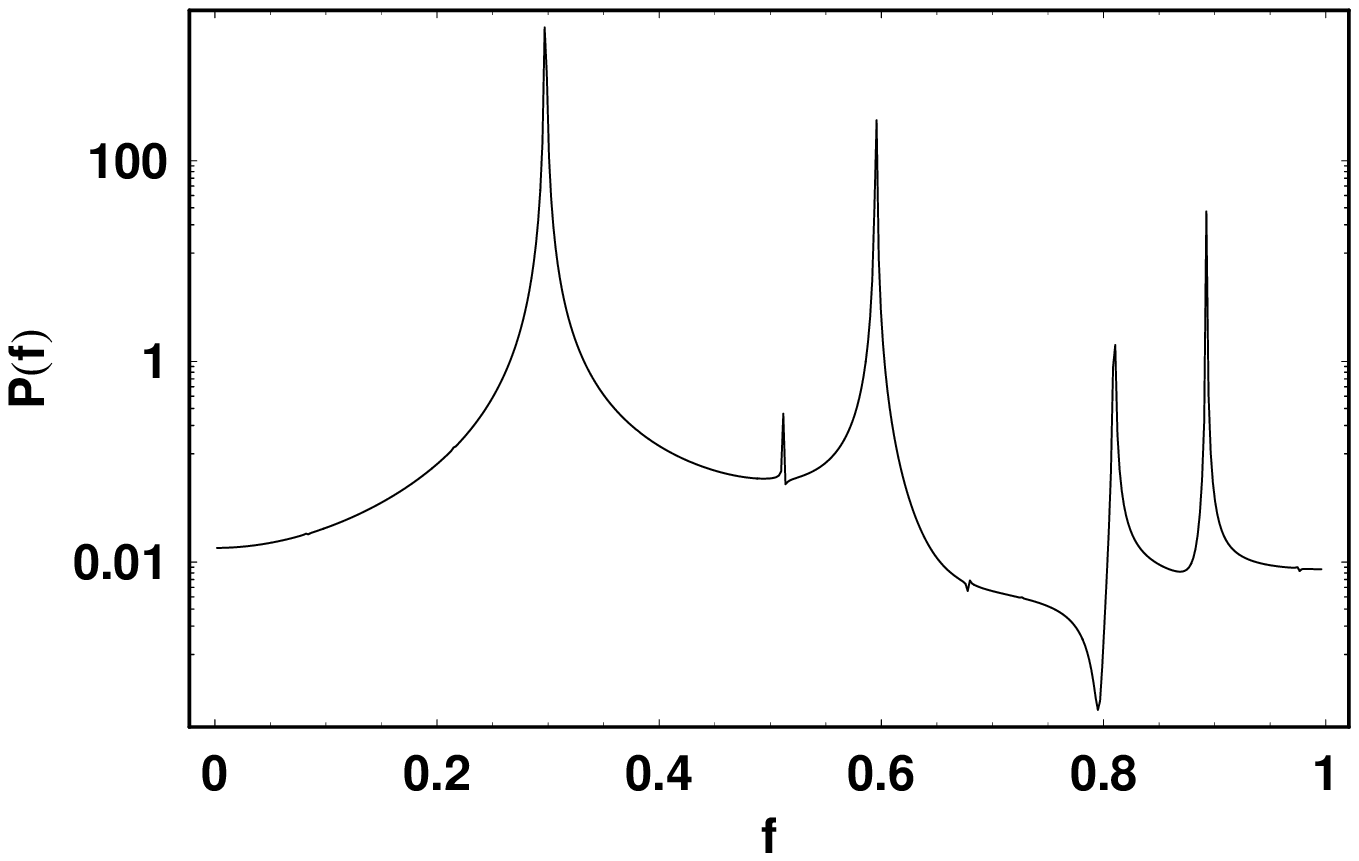}}}
\vskip 0.01cm
\caption{(a-d): (a-upper left): A tube orbit. The initial conditions are: $y_0=1.05, x_0 = p_{y0}=0$, (b-upper right): The evolution of the LCE, (c-down left): The $S(g)$ spectrum and (d-down right): The $P(f)$ indicator. The values of all the other parameters are as in Fig. 3.}
\end{figure*}
\begin{figure*}[!tH]
\centering
\resizebox{0.70\textwidth}{!}{\rotatebox{0}{\includegraphics*{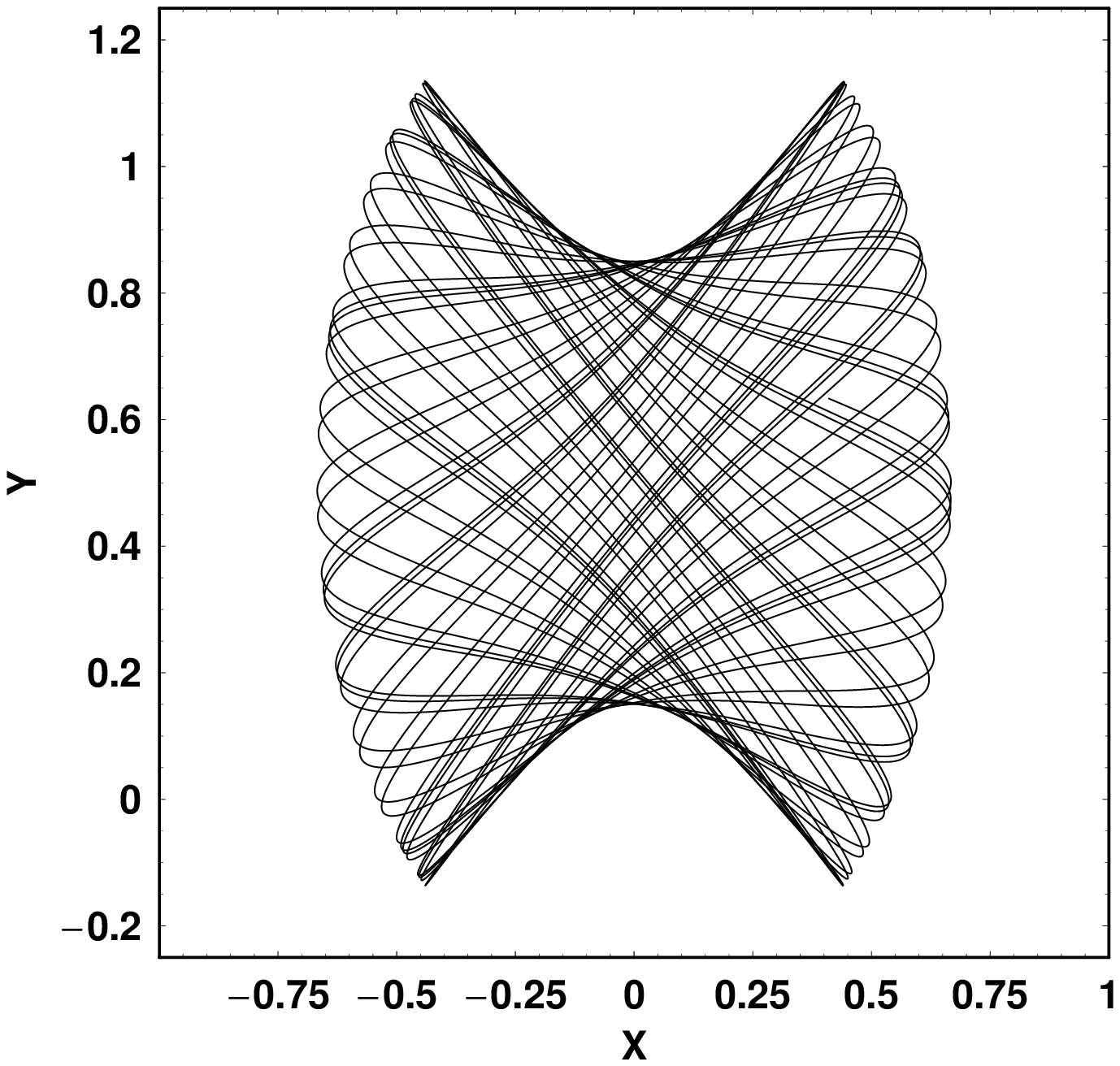}}\hspace{1cm}
                              \rotatebox{0}{\includegraphics*{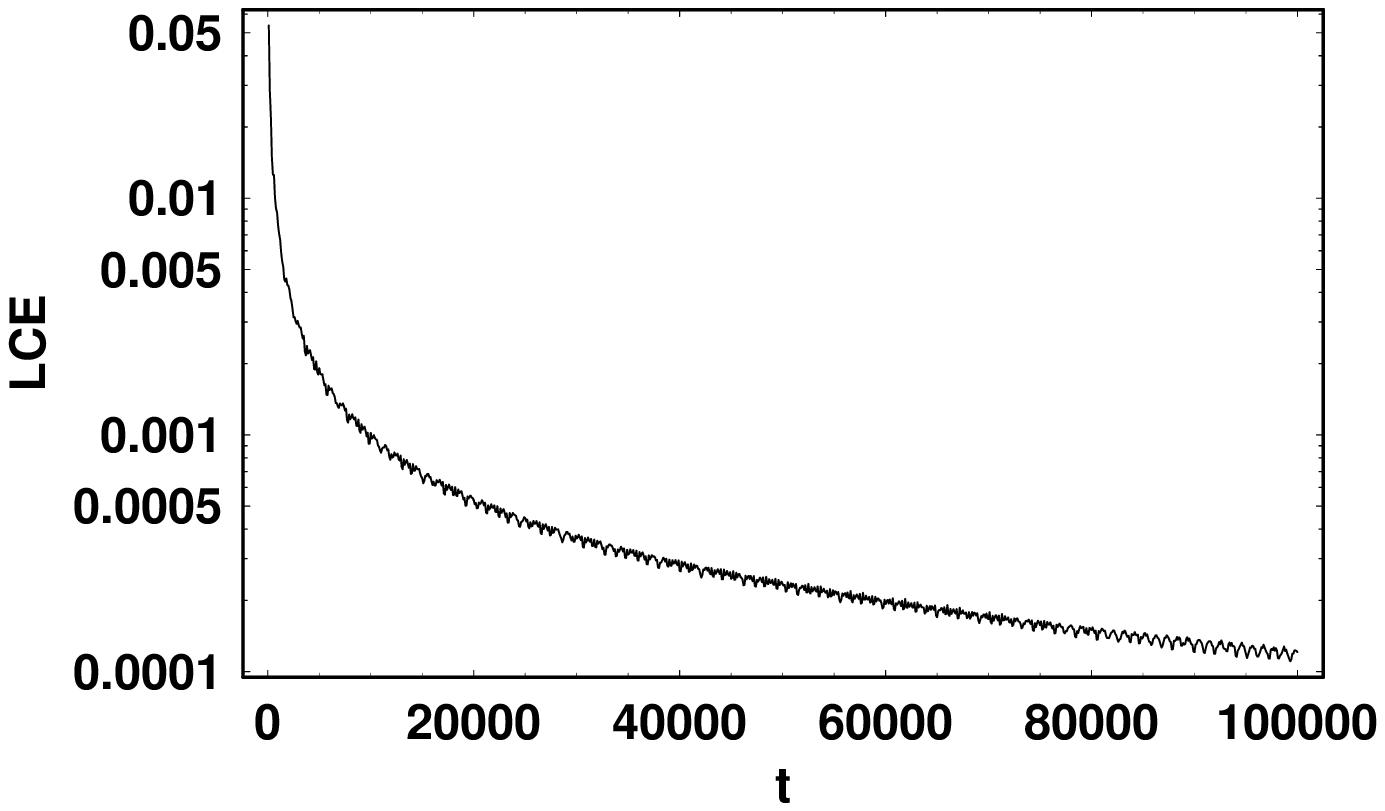}}}
\resizebox{0.70\textwidth}{!}{\rotatebox{0}{\includegraphics*{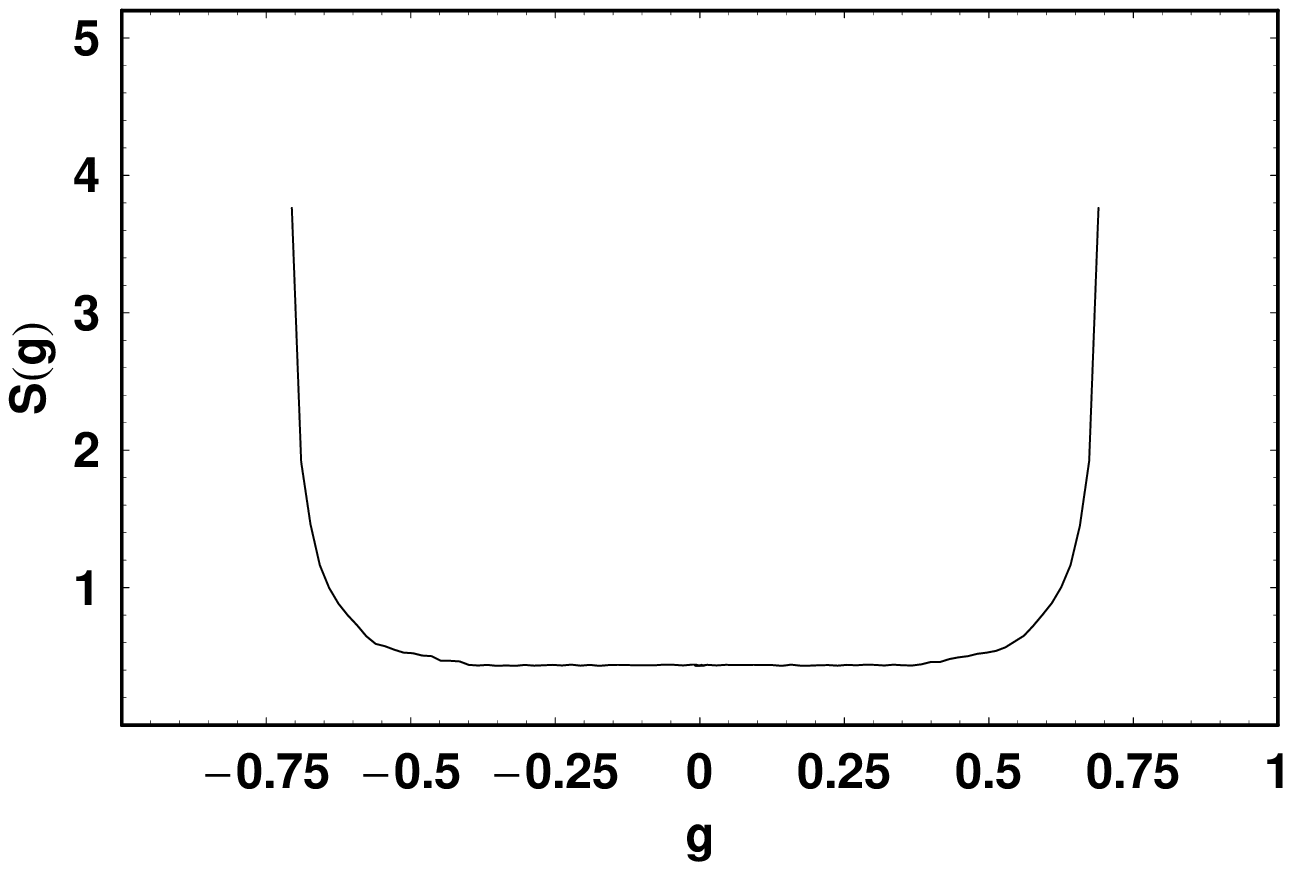}}\hspace{1cm}
                              \rotatebox{0}{\includegraphics*{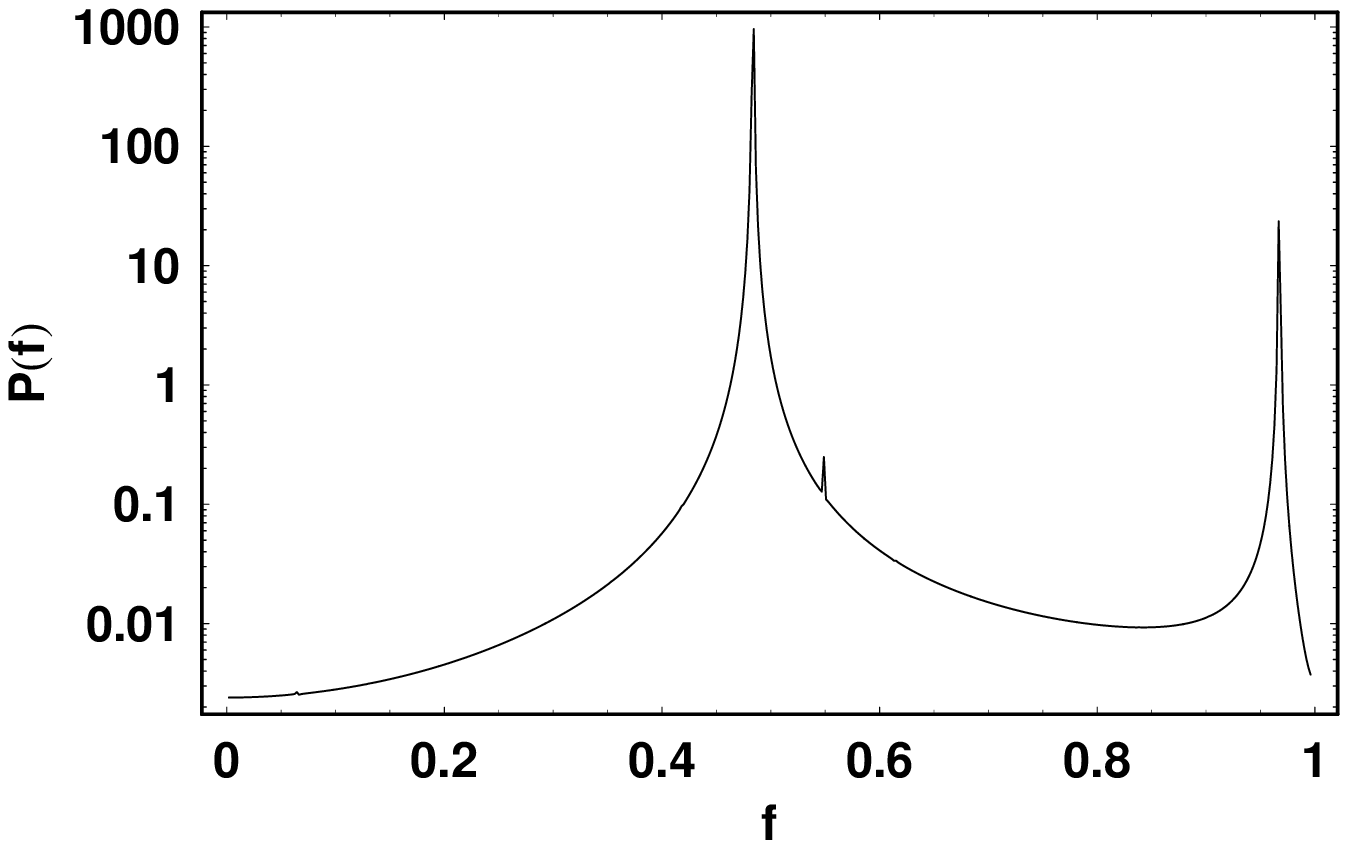}}}
\vskip 0.01cm
\caption{(a-d): (a-upper left): A box orbit. The initial conditions are: $y_0=0.85, x_0 = p_{y0}=0$, (b-upper right): The evolution of the LCE, (c-down left): The $S(g)$ spectrum and (d-down right): The $P(f)$ indicator. The values of all the other parameters are as in Fig. 3.}
\end{figure*}
\begin{figure}[!tH]
\centering
\resizebox{0.60\hsize}{!}{\rotatebox{270}{\includegraphics*{Fig-6.ps}}}
\caption{The $(y, p_y)$ phase plane when $A=4$, $B=-1$ and $c=0.24$. The value of the energy is $E_{el} = 0.3492$. $P1$ and $P2$ represent the two identical ellipses (3)traversed in opposite directions.}
\end{figure}
\begin{figure*}[!tH]
\centering
\resizebox{0.70\textwidth}{!}{\rotatebox{0}{\includegraphics*{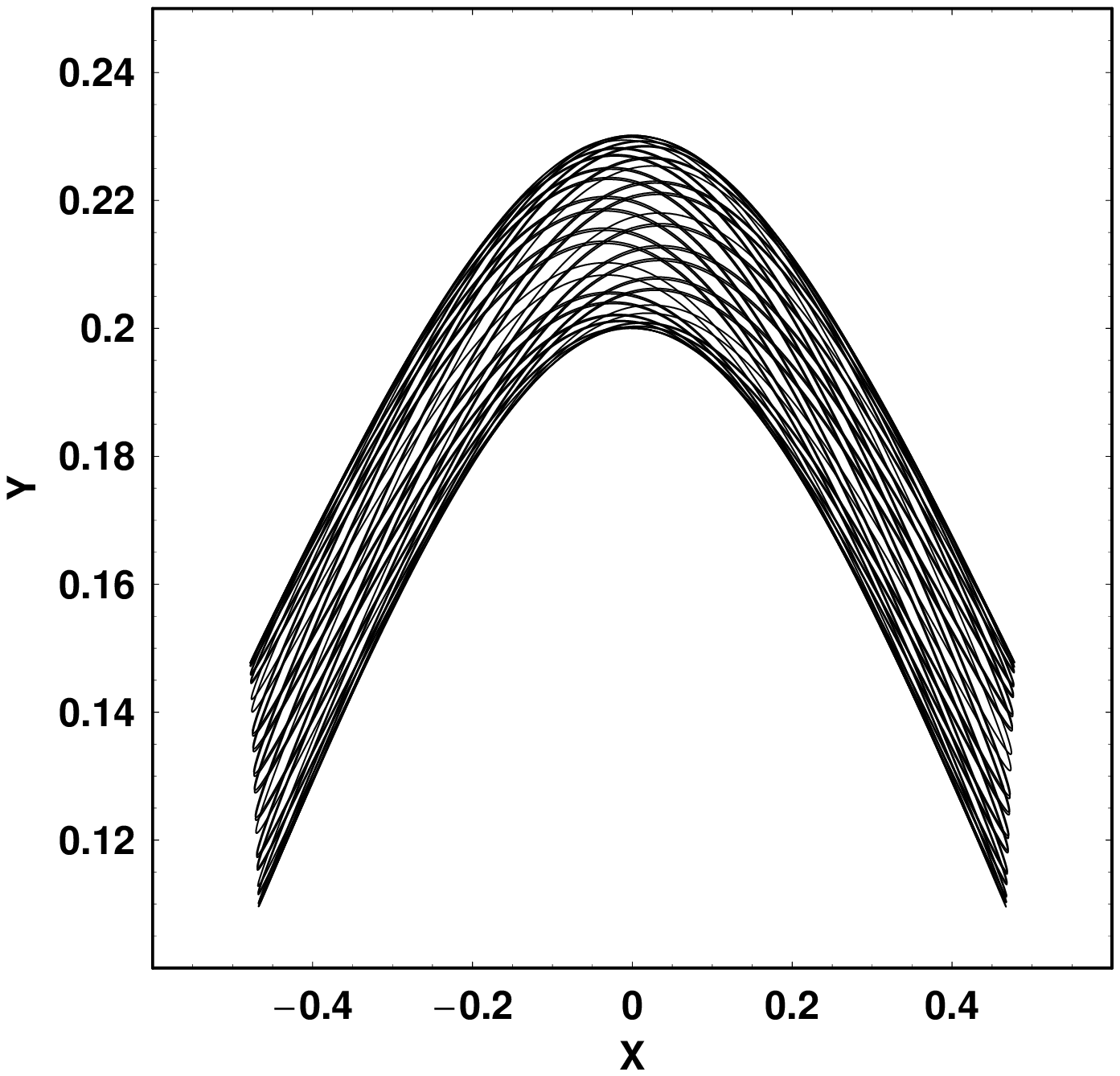}}\hspace{1cm}
                              \rotatebox{0}{\includegraphics*{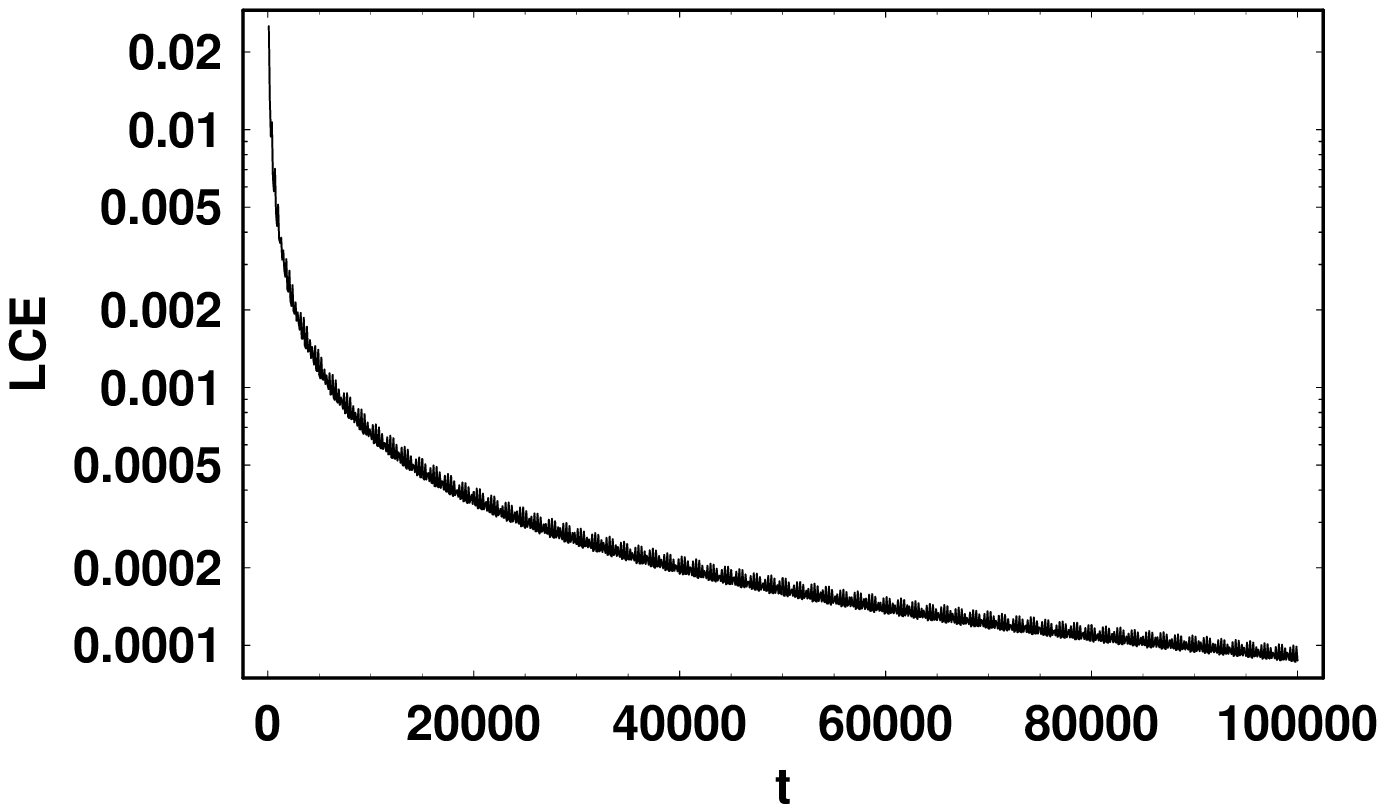}}}
\resizebox{0.70\textwidth}{!}{\rotatebox{0}{\includegraphics*{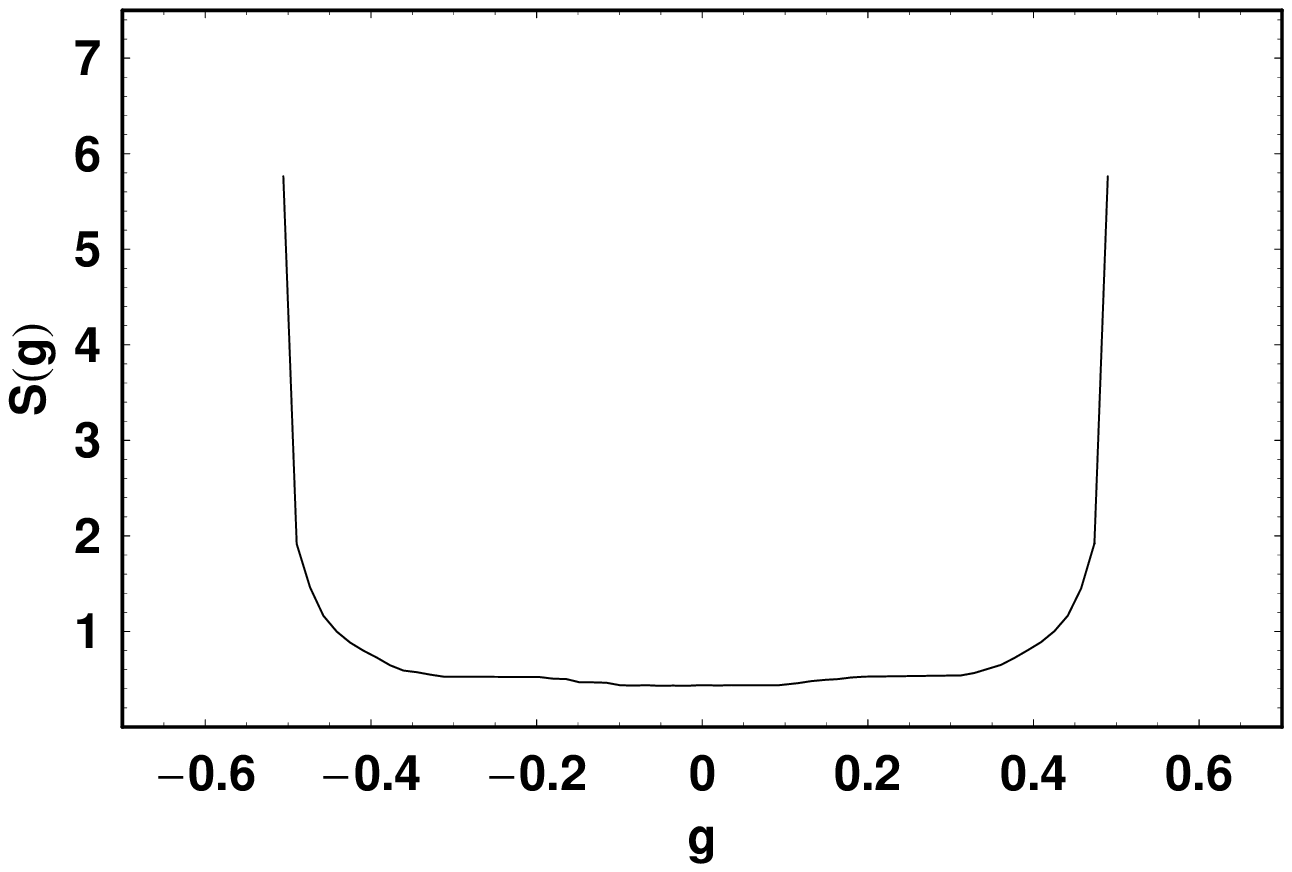}}\hspace{1cm}
                              \rotatebox{0}{\includegraphics*{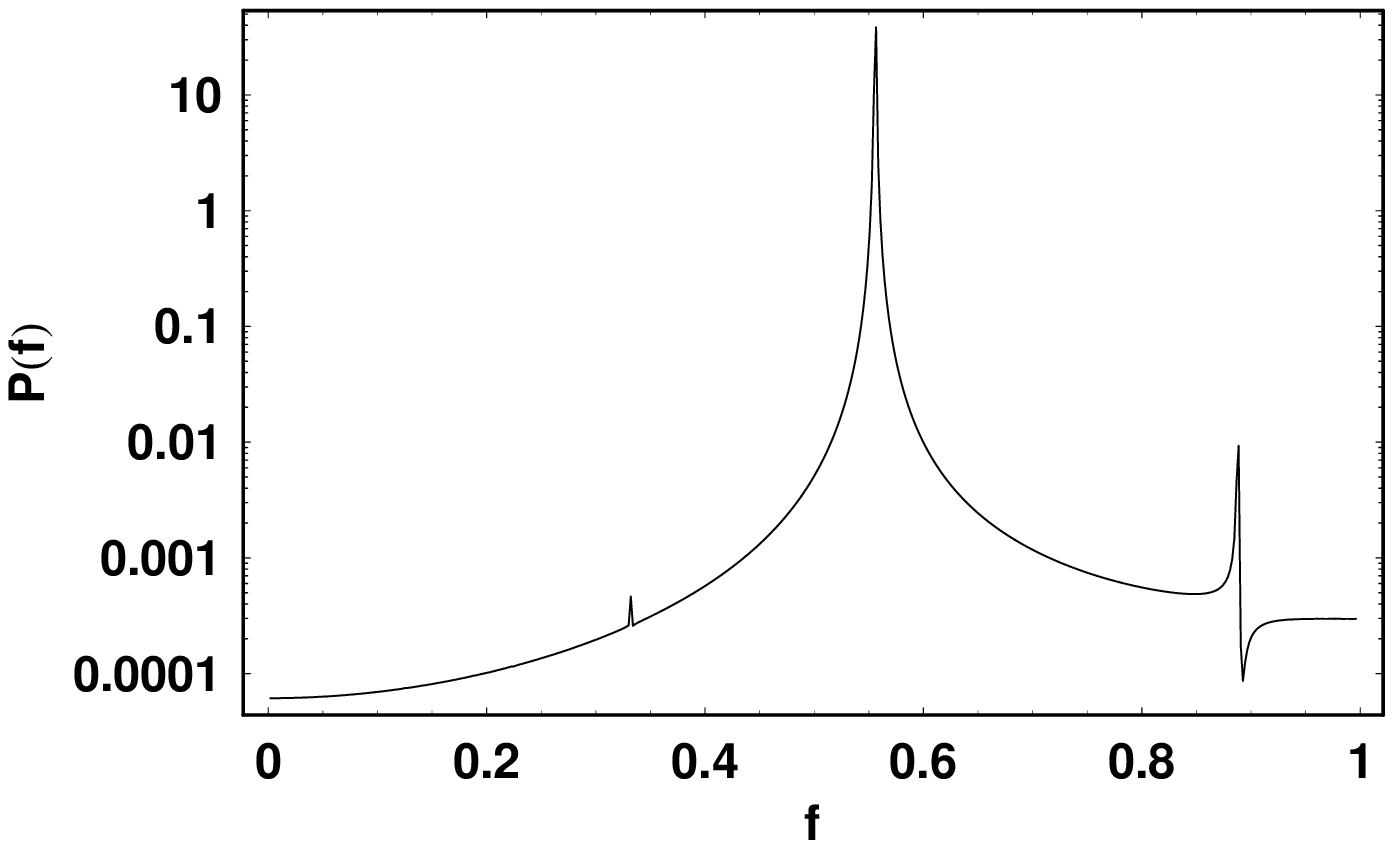}}}
\vskip 0.01cm
\caption{(a-d): (a-upper left): A resonant box orbit. The initial conditions are: $y_0=0.2, x_0 = p_{y0}=0$, (b-upper right): The evolution of the LCE, (c-down left): The $S(g)$ spectrum and (d-down right): The $P(f)$ indicator. The values of all the other parameters are as in Fig. 6.}
\end{figure*}
\begin{figure*}[!tH]
\centering
\resizebox{0.70\textwidth}{!}{\rotatebox{0}{\includegraphics*{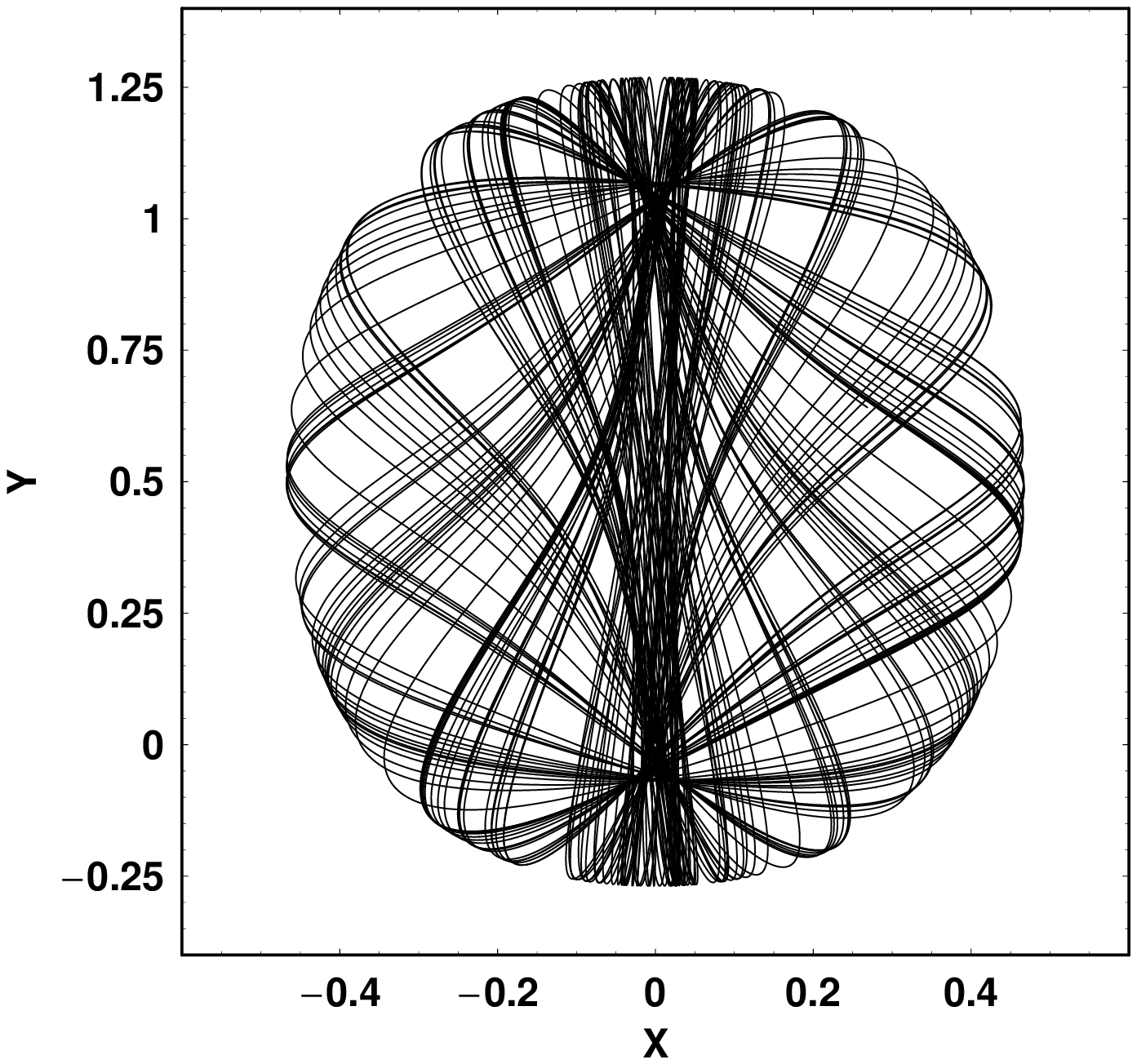}}\hspace{1cm}
                              \rotatebox{0}{\includegraphics*{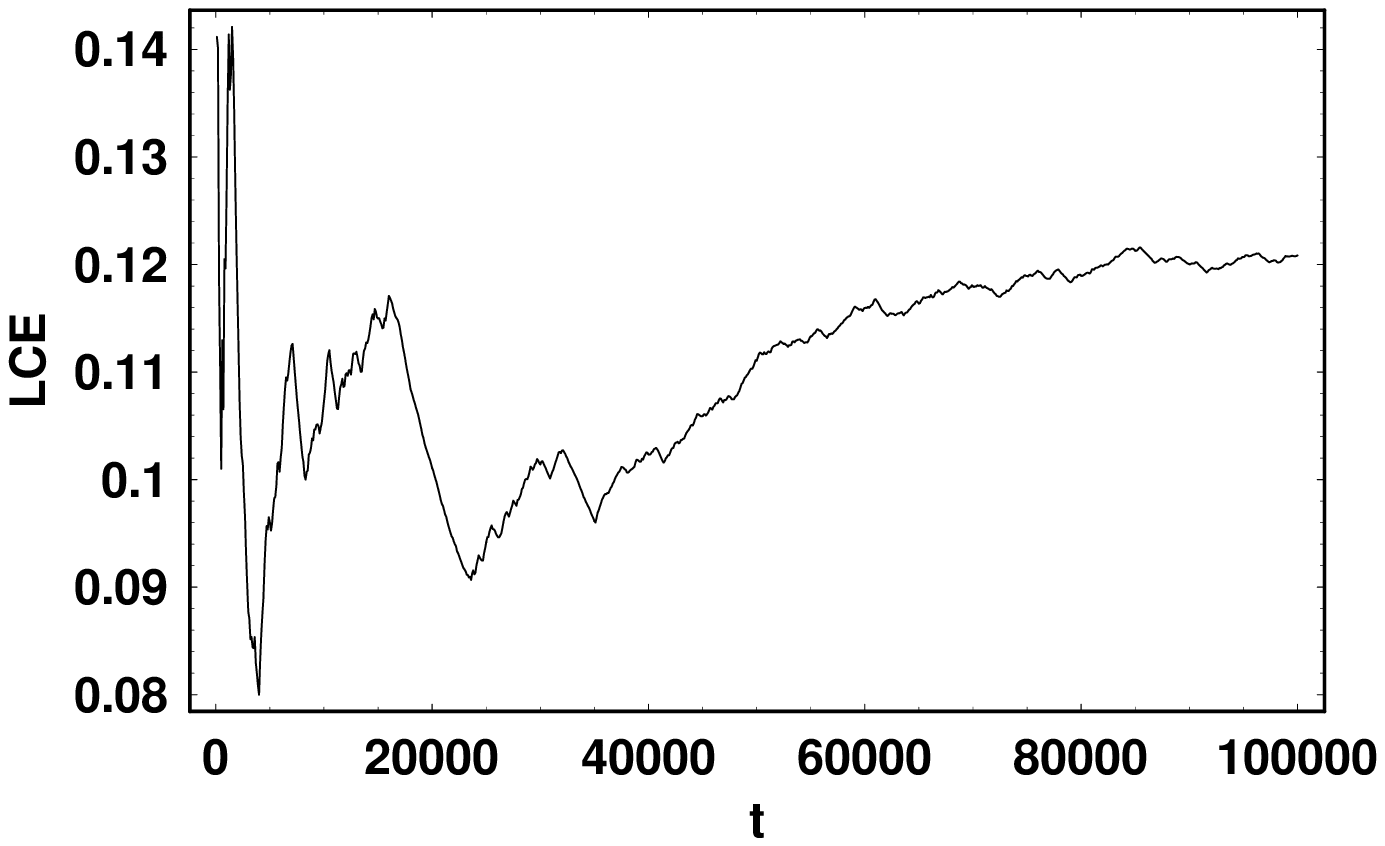}}}
\resizebox{0.70\textwidth}{!}{\rotatebox{0}{\includegraphics*{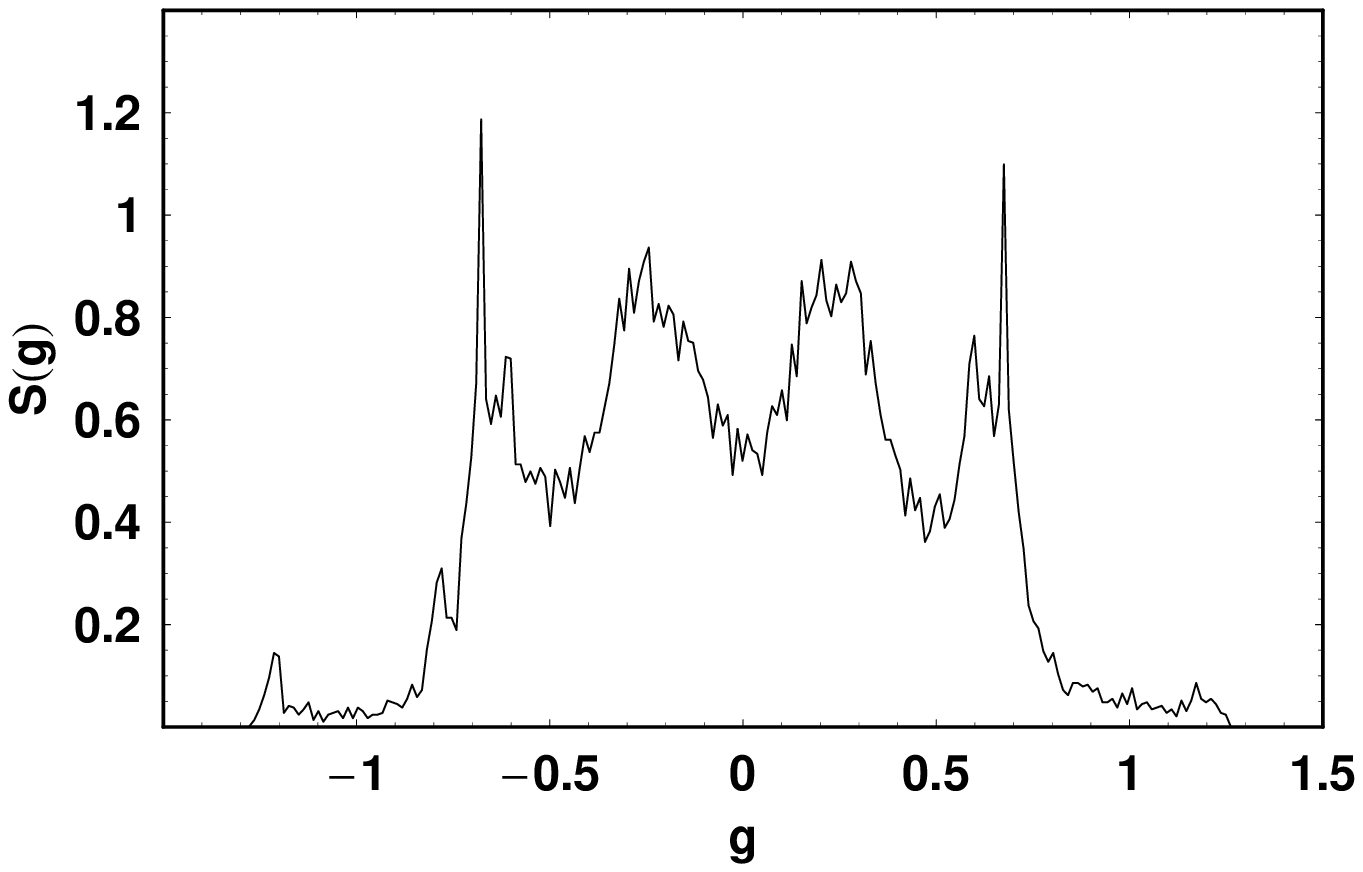}}\hspace{1cm}
                              \rotatebox{0}{\includegraphics*{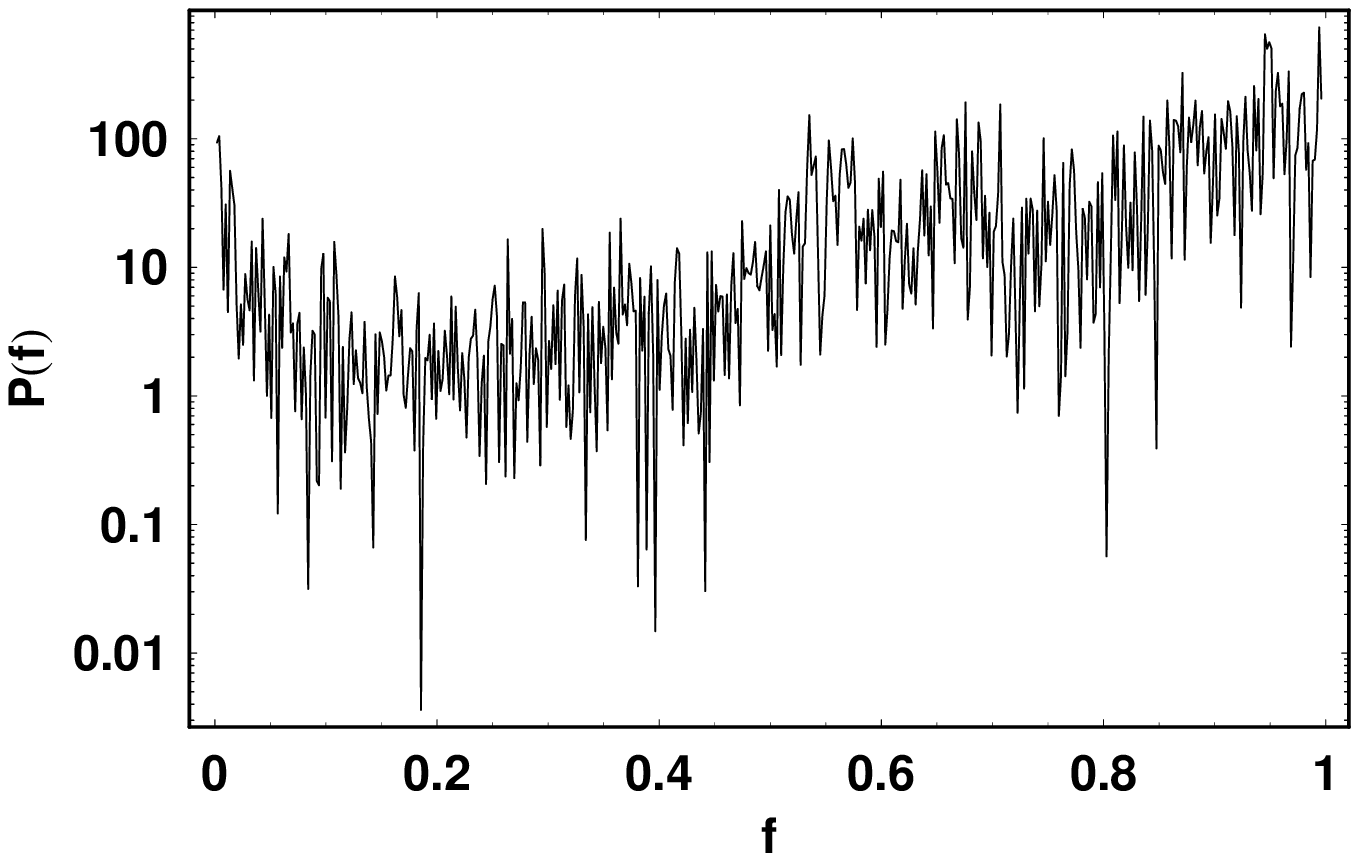}}}
\vskip 0.01cm
\caption{(a-d): (a-upper left): A chaotic orbit. The initial conditions are: $y_0=0.45, x_0 = p_{y0}=0$, ((b-upper right): The evolution of the LCE, (c-down left): The $S(g)$ spectrum and (d-down right): The $P(f)$ indicator. The values of all the other parameters are as in Fig. 6.}
\end{figure*}

Figure 6 shows the $(y, p_y)$ phase plane when $A=4$, $B=-1$ and $c=0.24$. The picture here is different from that shown in Fig. 3. The value of the energy is now $E_{el} = 0.3492$. The two ellipses are present and the corresponding tube orbits as well. In this case, the central invariant point is unstable and there are two more stable invariant points on the $y$-axis. The corresponding periodic orbits are oscillations near $L_1$. Orbits starting near the above two orbits are resonant box orbits. These resonant box orbits, form a bulge like structure in the central parts of the bar. Such a resonant box orbit is shown in
Figure 7a, while in Figure 7b we can observe the time evolution of the LCE. Figure 7c shows the $S(g)$ spectrum, while Figure 7d the corresponding $P(f)$ indicator. The initial conditions are: $y_0=0.2, x_0 = p_{y0}=0$. What is more interesting is the considerable chaotic region surrounding the invariant curves produced by the resonant box orbits. This chaotic region is produced by orbits starting near the central unstable periodic point. A chaotic orbit is shown in Figure 8a. The initial conditions are: $y_0=0.45, x_0 = p_{y0}=0$, while in Figure 8b we see a typical evolution of the LCE of a chaotic orbit. In Figure 8c, one can observe that the $S(g)$ spectrum is highly asymmetric with a large number of small and large peaks, which is clearly an indication of chaotic motion. Finally, in Figure 8d we can see a similar profile, regarding the $P(f)$ indicator. All three indicators support the chaotic character of this orbit. As one can see, the chaotic orbits support the barred structure.
\begin{figure*}[!tH]
\centering
\resizebox{0.90\textwidth}{!}{\rotatebox{0}{\includegraphics*{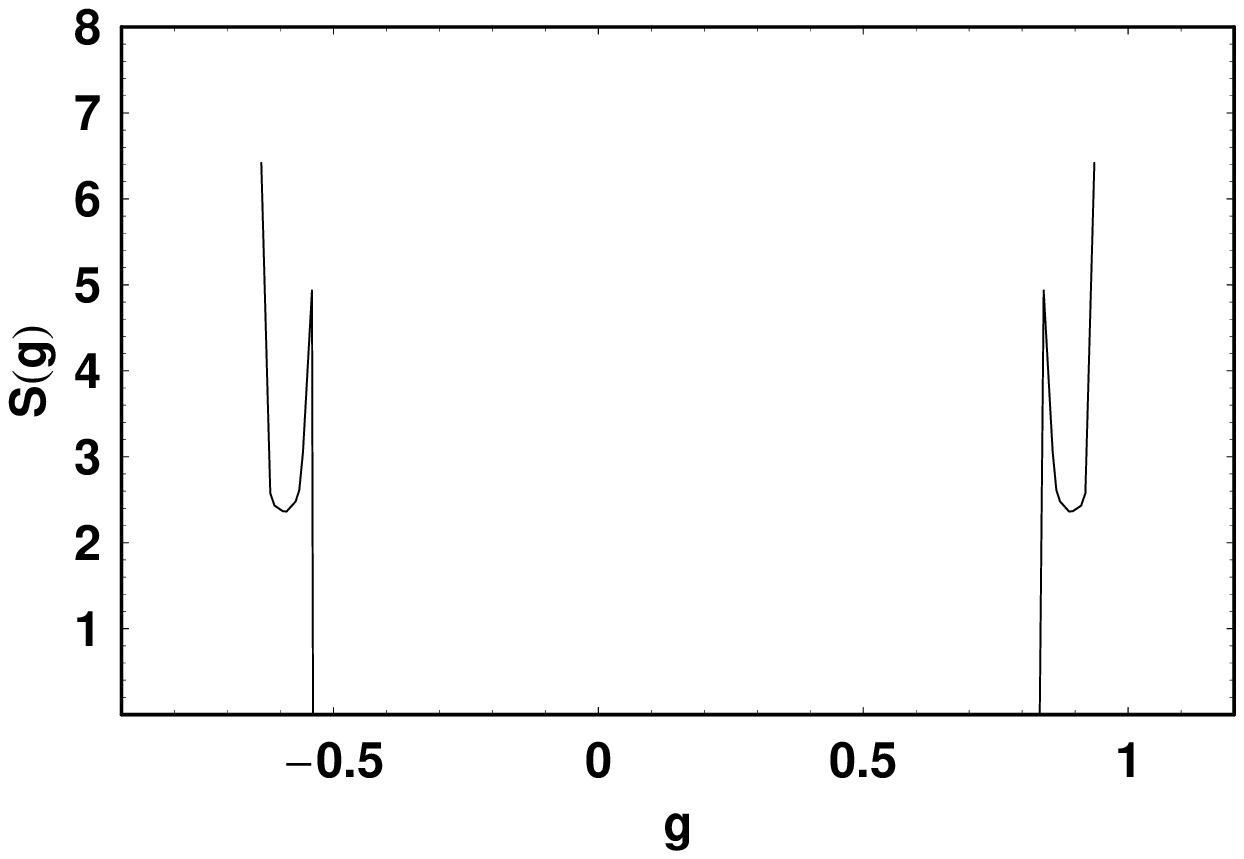}}\hspace{1cm}
                              \rotatebox{0}{\includegraphics*{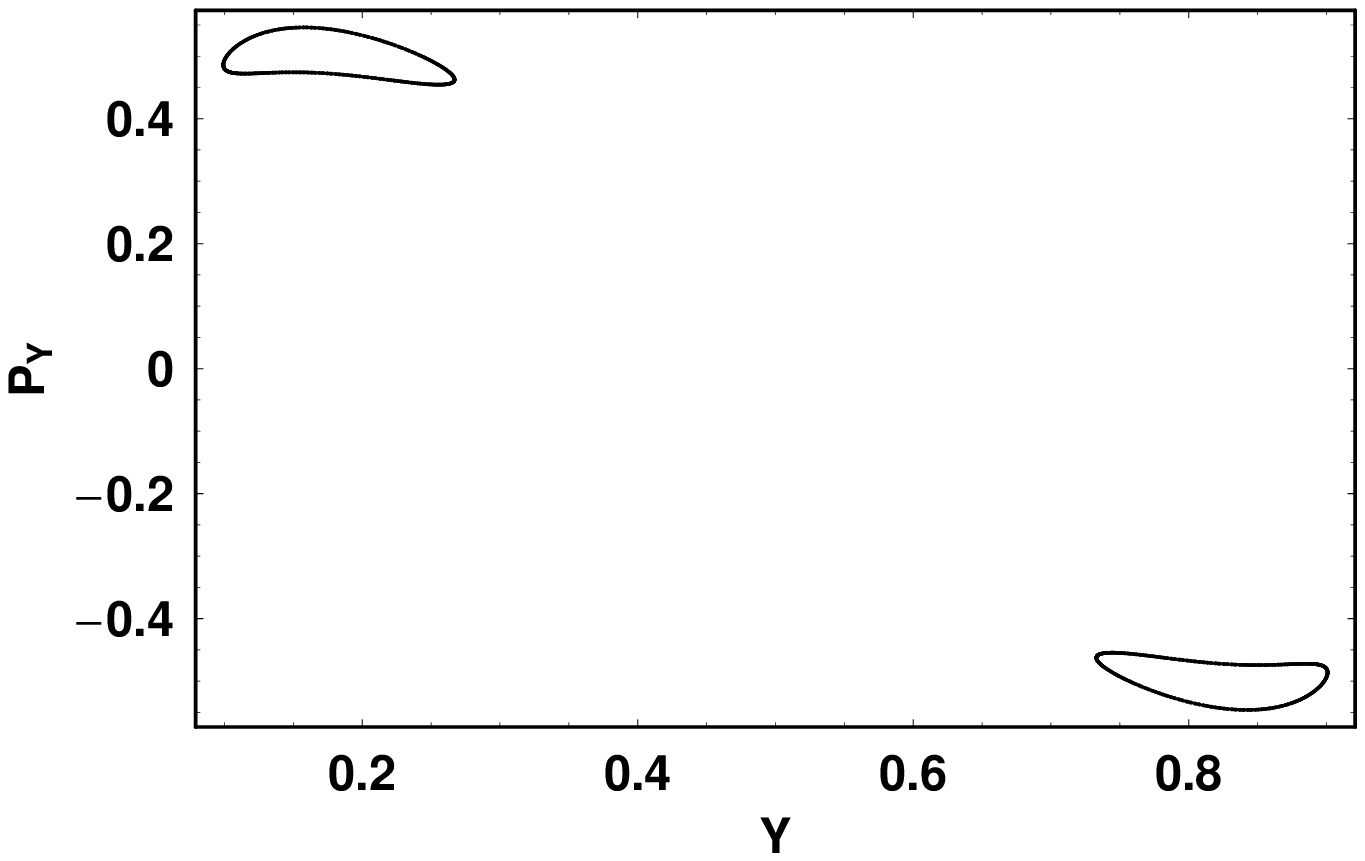}}}
\vskip 0.01cm
\caption{(a-b): (a-left): The $S(g)$ spectrum for a resonant orbit producing two small islands of invariant curves. (b-right): The two small islands on the $(y, p_y)$ surface of section of Fig. 6.}
\end{figure*}
\begin{figure*}[!tH]
\centering
\resizebox{0.90\textwidth}{!}{\rotatebox{0}{\includegraphics*{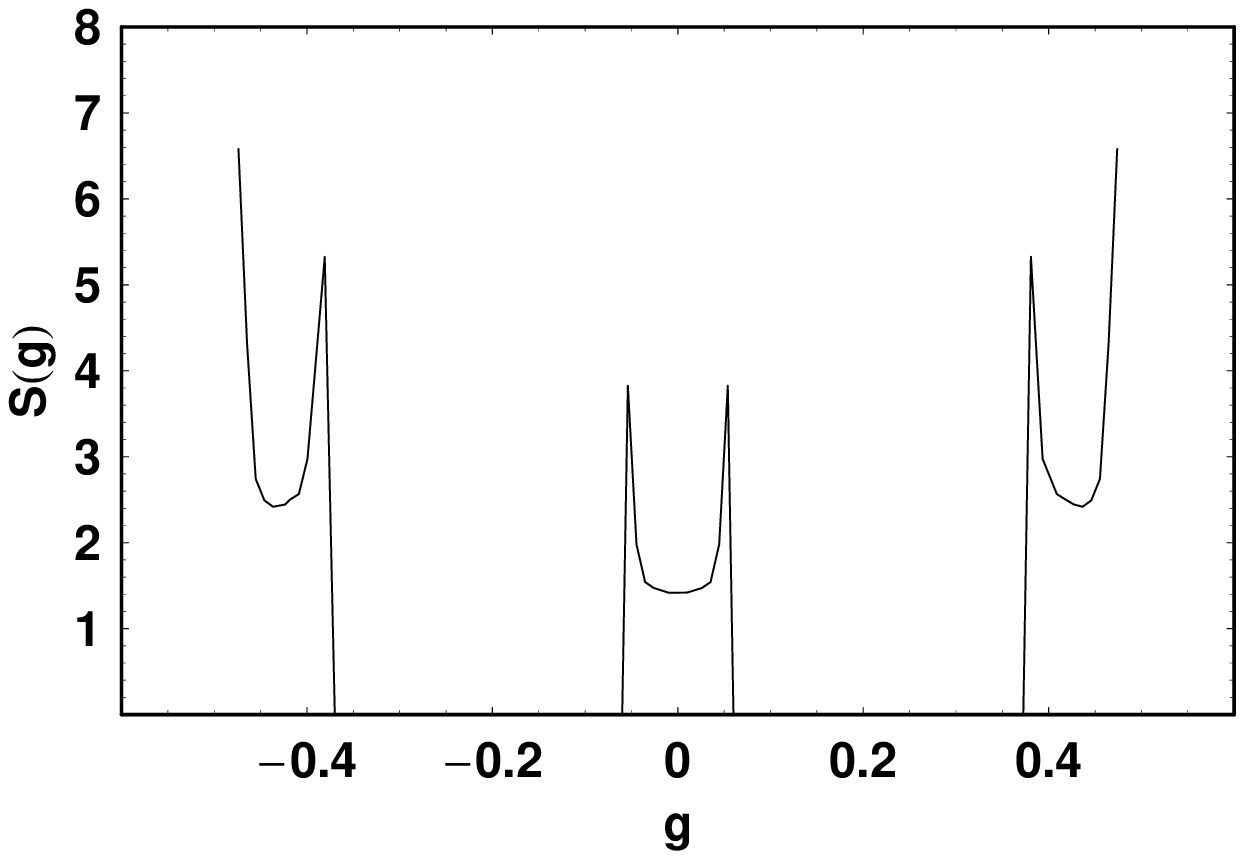}}\hspace{1cm}
                              \rotatebox{0}{\includegraphics*{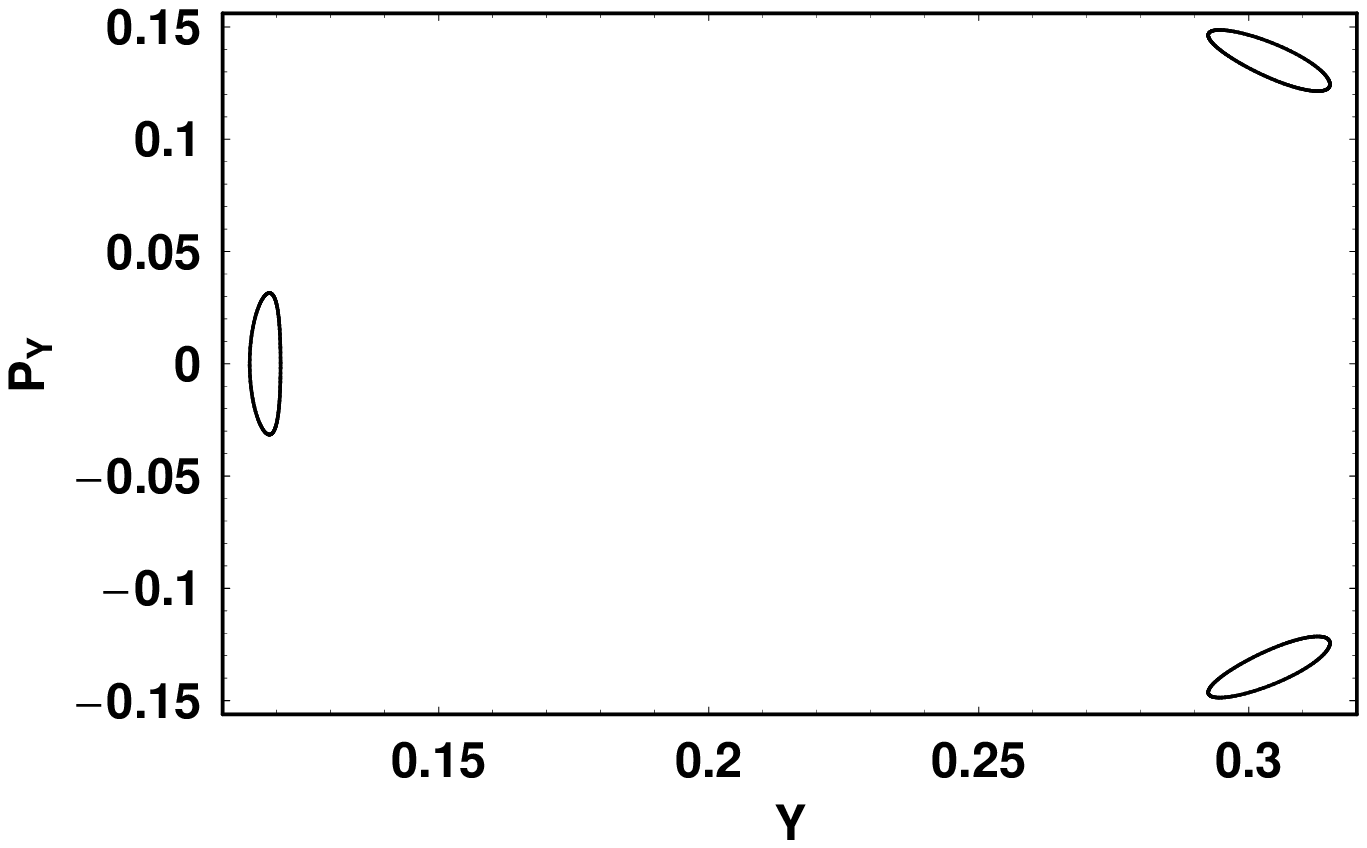}}}
\vskip 0.01cm
\caption{(a-b): (a-left): The $S(g)$ spectrum for a resonant orbit producing three small islands of invariant curves. (b-right): The three small islands on the $(y, p_y)$ surface of section of Fig. 6.}
\end{figure*}
\begin{figure*}[!tH]
\centering
\resizebox{0.90\textwidth}{!}{\rotatebox{0}{\includegraphics*{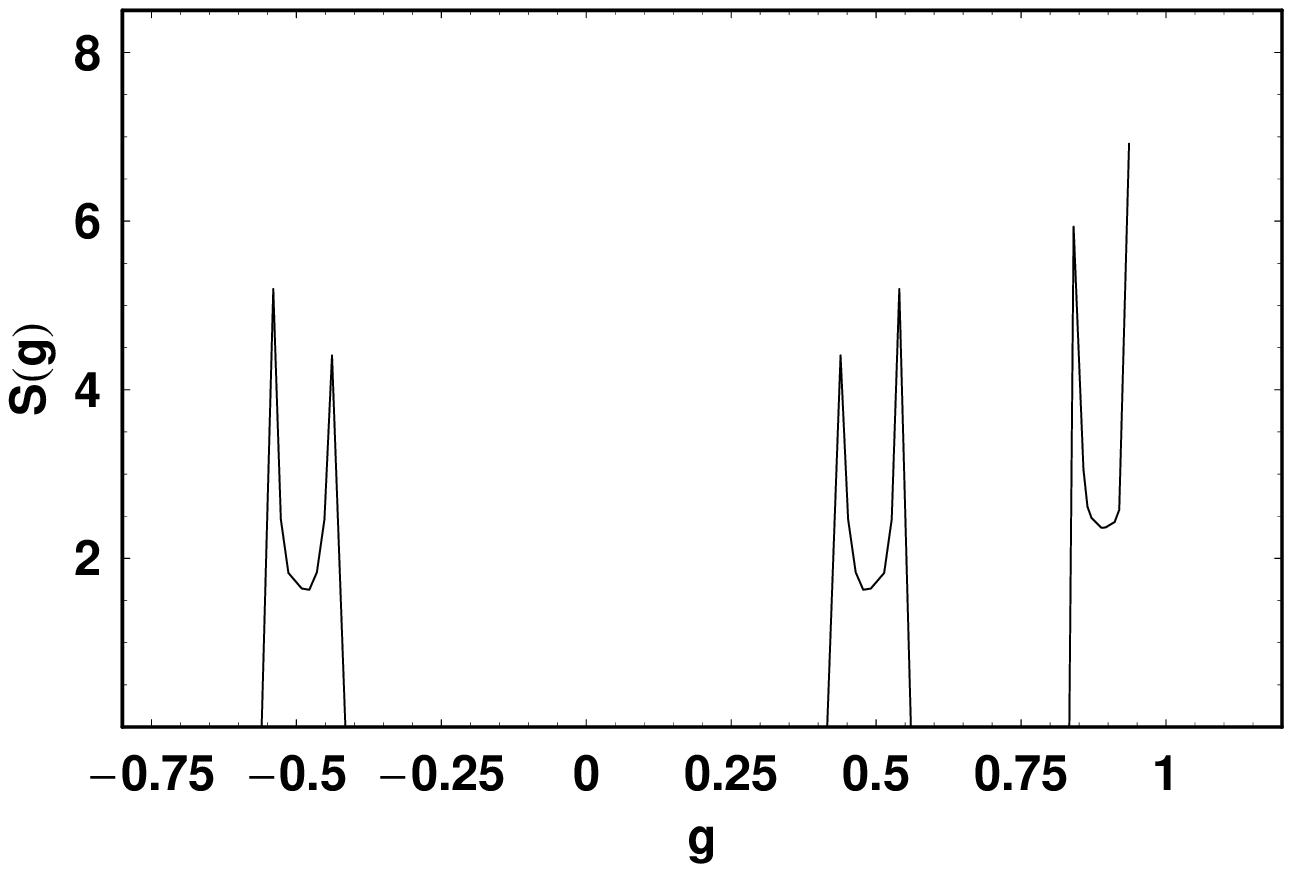}}\hspace{1cm}
                              \rotatebox{0}{\includegraphics*{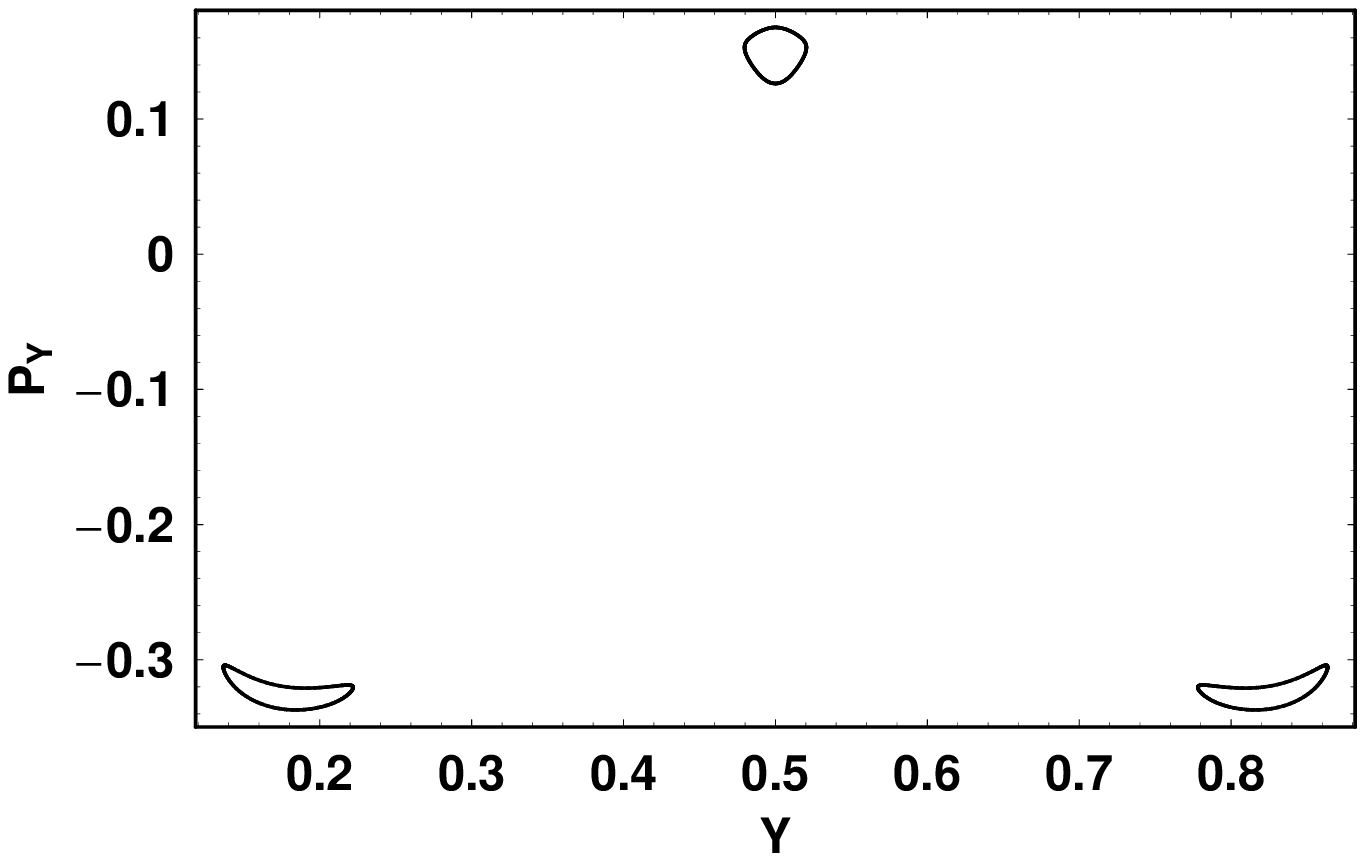}}}
\vskip 0.01cm
\caption{(a-b): (a-left): The $S(g)$ spectrum for an orbit producing a set of three small islands of invariant curves. (b-right): The set of the three small islands on the $(y, p_y)$ surface of section of Fig. 6.}
\end{figure*}

Figure 9a shows the $S(g)$ spectrum for a resonant orbit producing two islands of invariant curves shown in Figure 9b. The initial conditions are: $y_0=0.1, x_0=0, p_{y0}=0.48$, while the value of $p_{x0}$ is always found from the energy integral. Figure 10a is similar to Fig. 9a and depicts the $S(g)$ spectrum for a resonant orbit producing three small islands shown in Figure 10b. The initial conditions are: $y_0=0.115, x_0=p_{y0}=0$. Here, we observes three $U$-type spectra, that is as much as the number of islands of invariant curves. Furthermore, we see that the left and right spectra are nearly symmetrical about the $g = 0$ axis, while the central spectrum lies on both sides of this axis. This indicates that, two of the islands are symmetric about the $y$-axis and the third intersects the $y-$axis. In other words, we have a quasi periodic orbit with a starting point on the $y-$axis. In Figure 11a-b we observe a resonant orbit producing a set of three small islands of invariant curves and three well defined $U$-type spectra. The initial conditions are: $y_0=0.48, x_0=0, p_{y0}=0.15$. The values of all other parameters are as in Fig. 6.  Thus, we may conclude that the $S(g)$ spectrum is a very useful dynamical indicator in order to identify islandic motion of resonant orbits. Here, we must emphasize that, with some previously used spectra (see Karanis \& Caranicolas, 2002 and references therein), we were able to detect islandic motion, but the number of spectra in some cases was smaller than the number of islands, because symmetric islands produced identical spectra. With the improved $S(g)$ spectrum, we have managed to get through this drawback.
\begin{figure*}[!tH]
\centering
\resizebox{0.90\textwidth}{!}{\rotatebox{0}{\includegraphics*{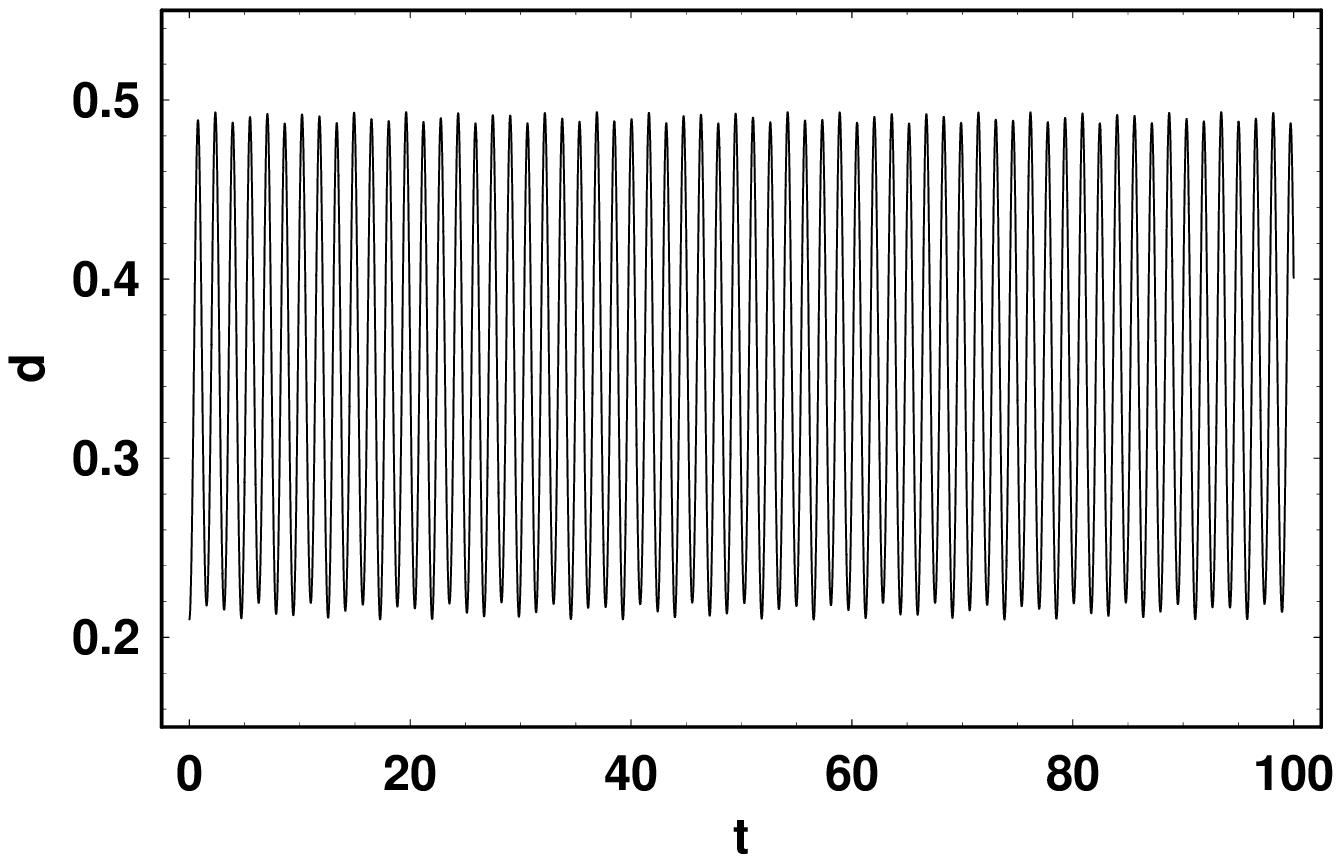}}\hspace{1cm}
                              \rotatebox{0}{\includegraphics*{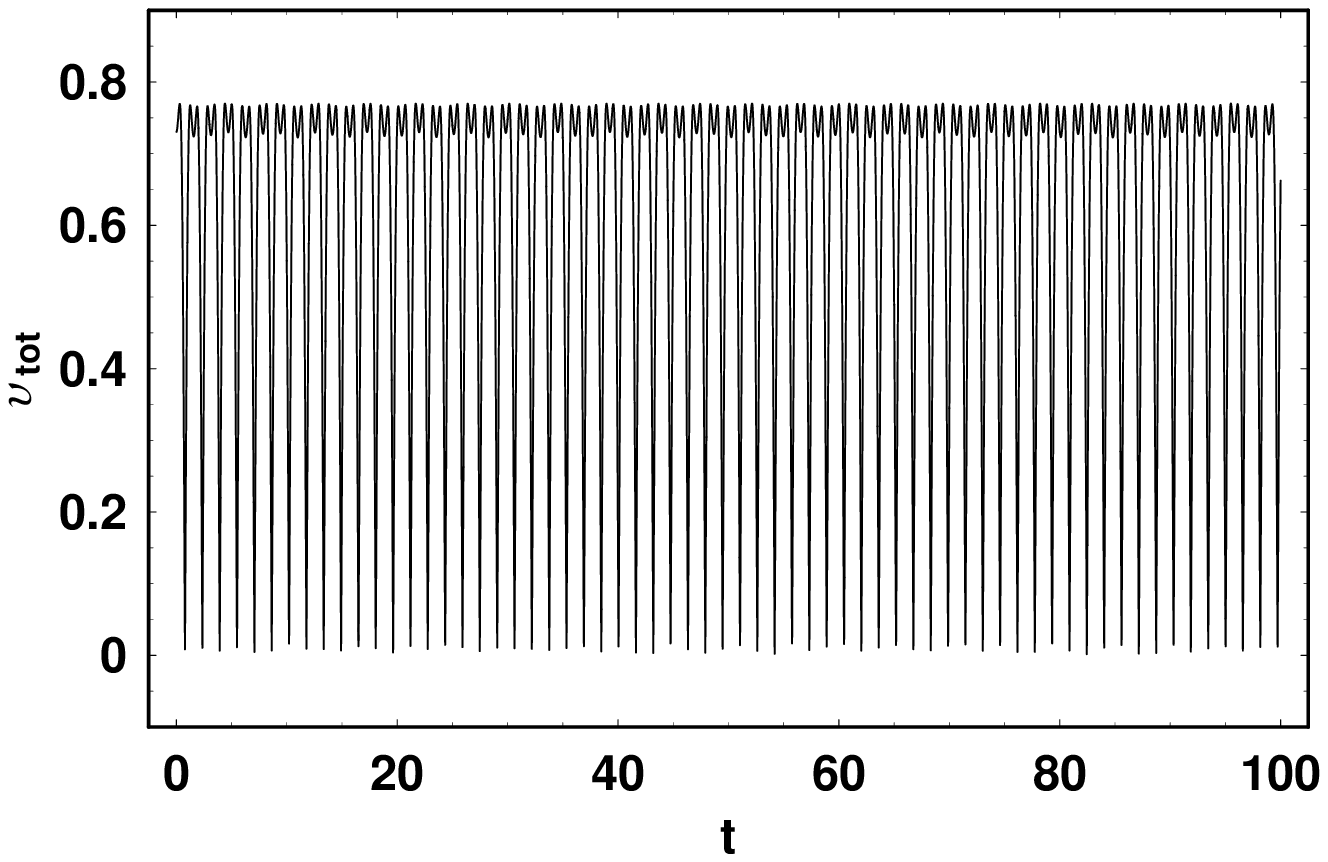}}}
\resizebox{0.90\textwidth}{!}{\rotatebox{0}{\includegraphics*{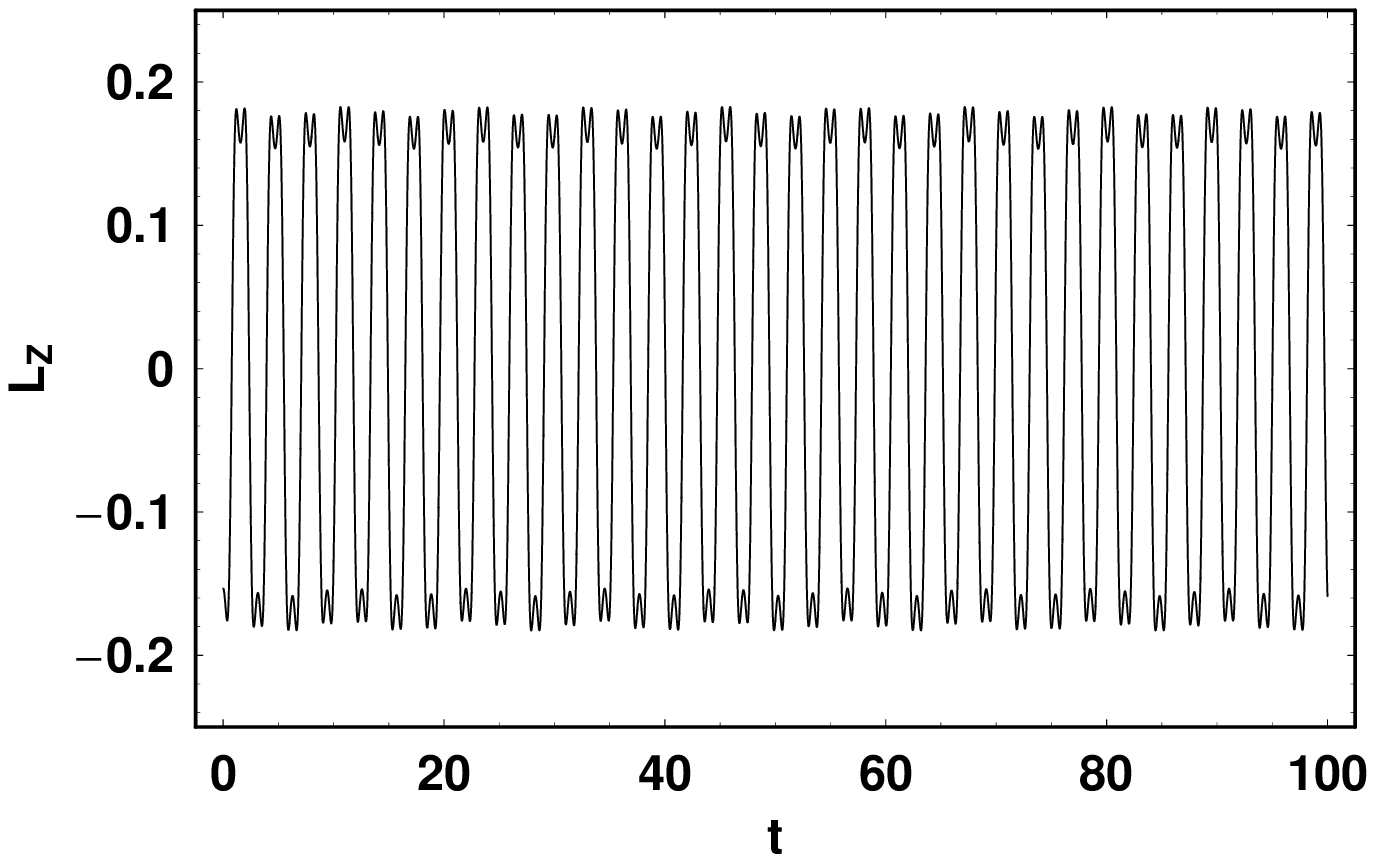}}\hspace{1cm}
                              \rotatebox{0}{\includegraphics*{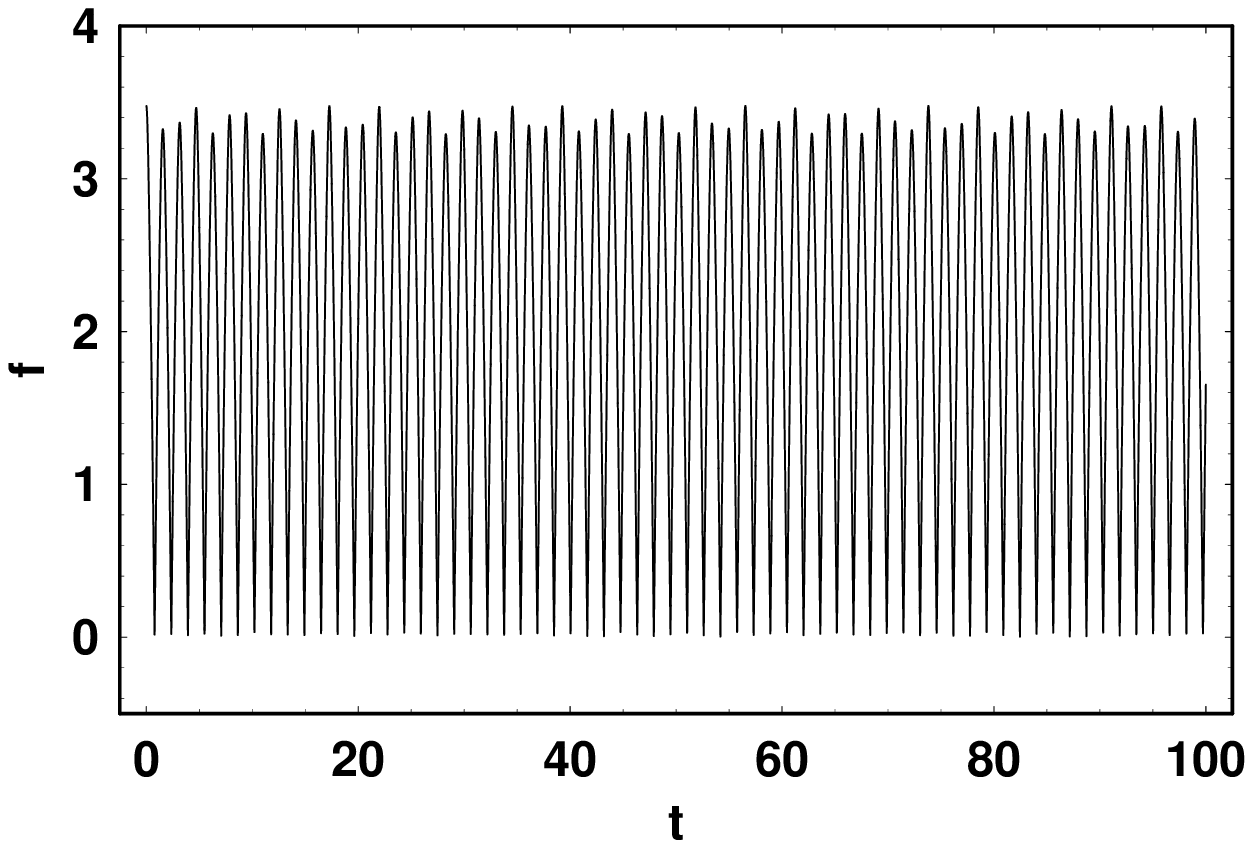}}}
\vskip 0.01cm
\caption{(a-d): Four dynamical indicators for the regular orbit of Fig. 7a. (a-upper left): distance versus time, (b-upper right): velocity versus time, (c-down left): $L_z$ versus time and (d-down right): orbital frequency versus time.}
\end{figure*}
\begin{figure*}[!tH]
\centering
\resizebox{0.90\textwidth}{!}{\rotatebox{0}{\includegraphics*{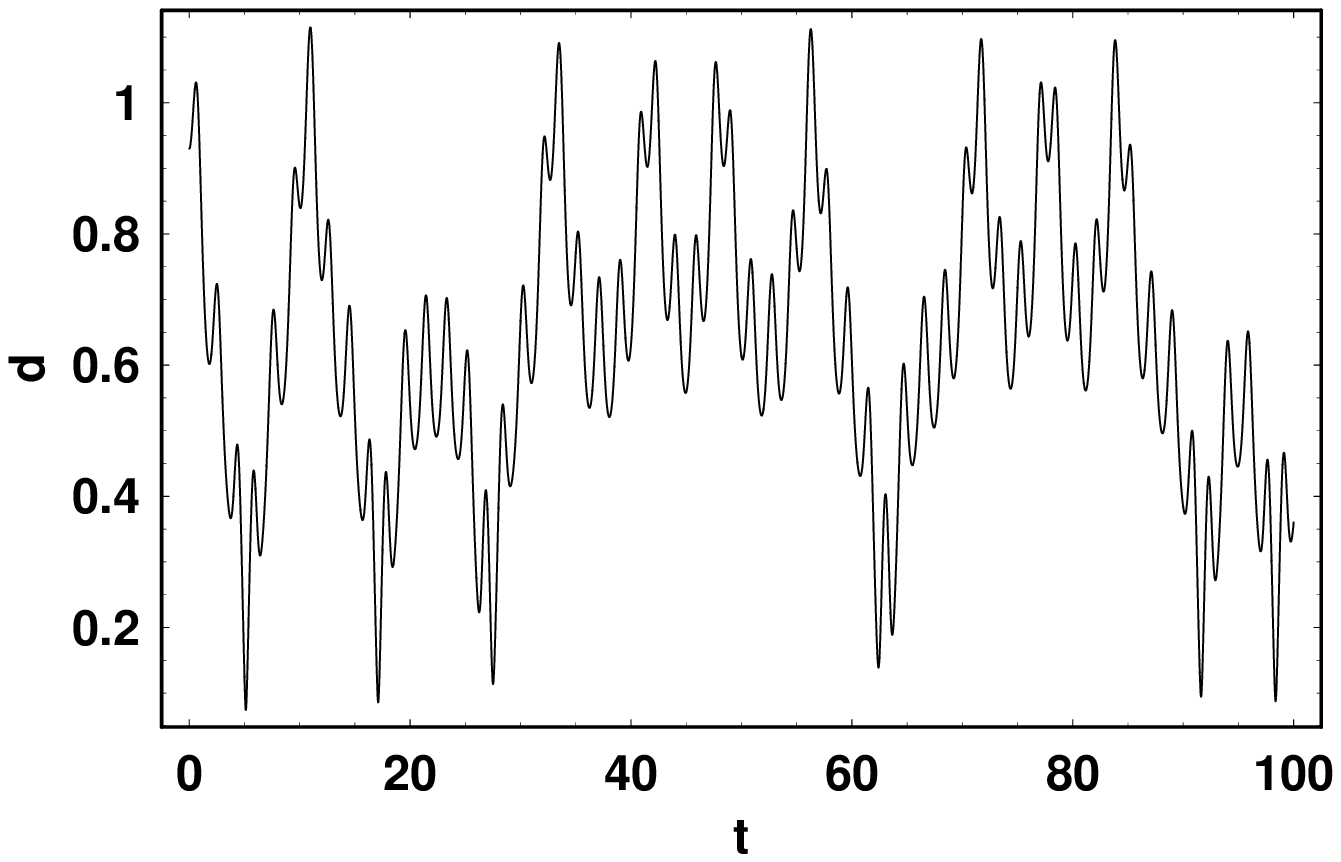}}\hspace{1cm}
                              \rotatebox{0}{\includegraphics*{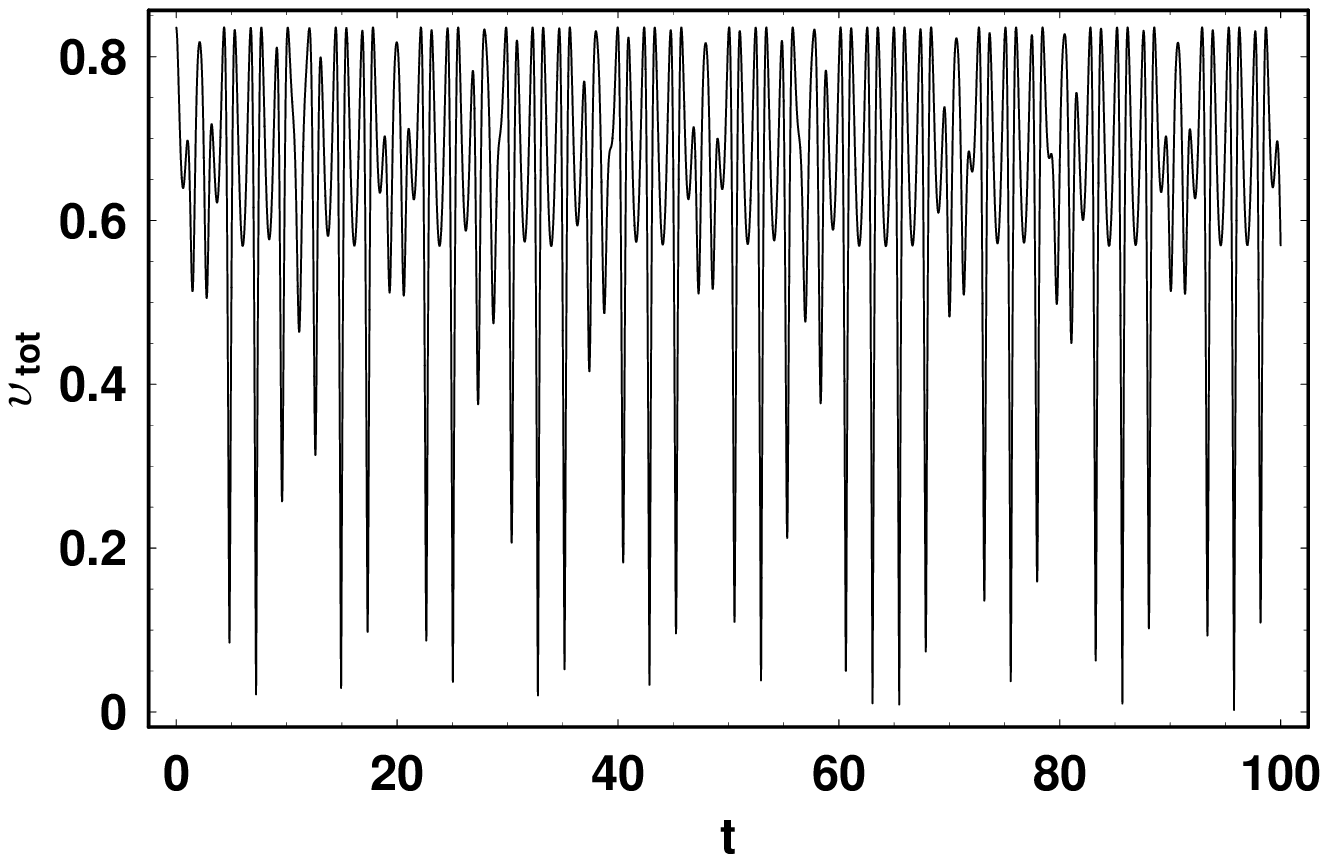}}}
\resizebox{0.90\textwidth}{!}{\rotatebox{0}{\includegraphics*{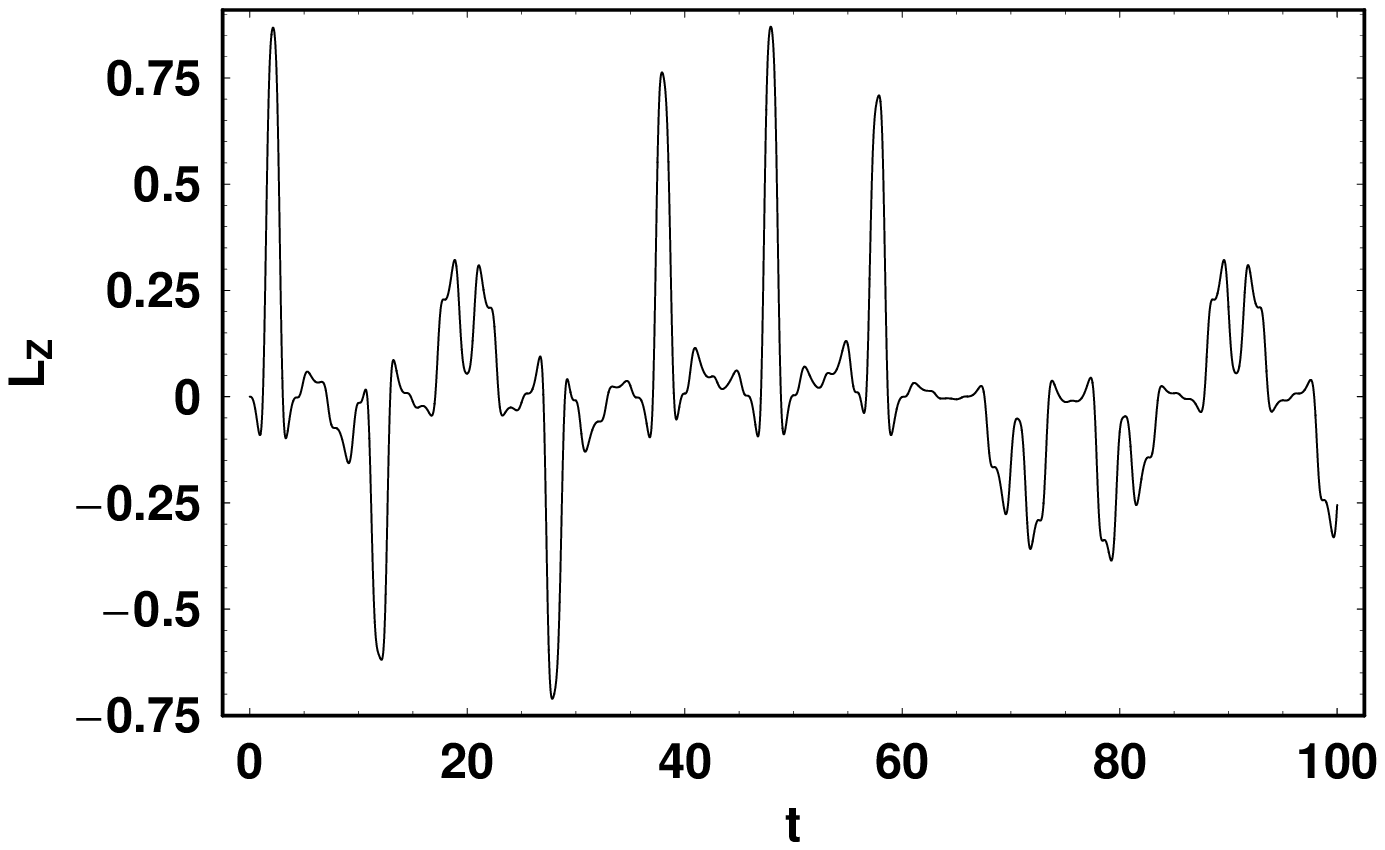}}\hspace{1cm}
                              \rotatebox{0}{\includegraphics*{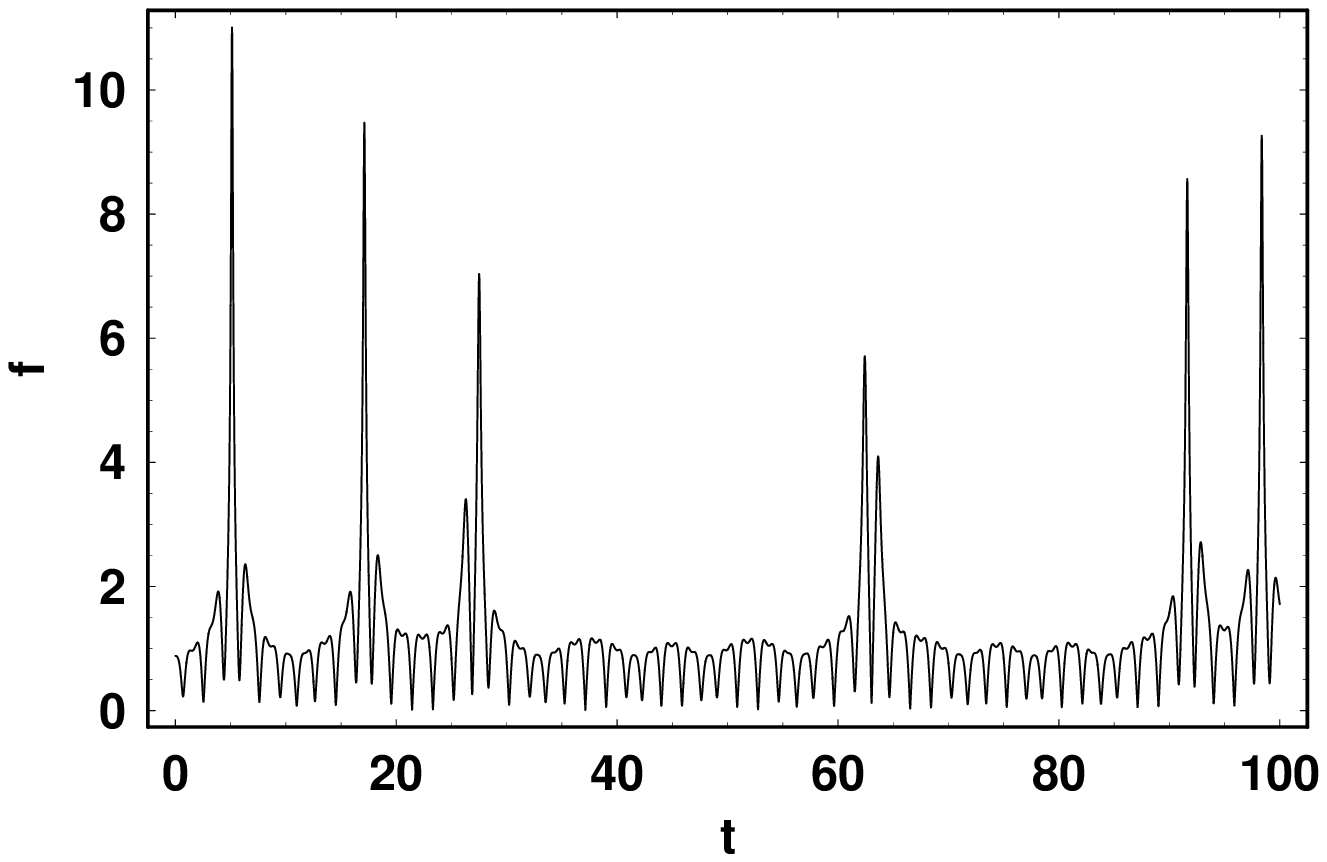}}}
\vskip 0.01cm
\caption{(a-d): Four dynamical indicators for the chaotic orbit of Fig. 8a. (a-upper left): distance versus time, (b-upper right): velocity versus time, (c-down left): $L_z$ versus time and (d-down right): orbital frequency versus time.}
\end{figure*}

At this point, we shall introduce some fast dynamical indicators, which they can help us to distinguish between order and chaos in galactic potentials. First we can see the time evolution of the profile of the distance $d = \sqrt{x^2 + y^2}$ of a test-particle (star) from the center of the galaxy. In the same way, we can observe the profile of the total velocity $\upsilon _{tot} = \sqrt{\upsilon _x^2 + \upsilon _y^2}$ versus time. Moreover, we can use a physical parameter, which plays an important role to the dynamical system, which is the angular momentum $L_z = xp_y - yp_x$. From previous experience, we know that low angular momentum stars, on approaching a dense and massive nucleus are scattered off the galactic plane displaying chaotic motion (see Caranicolas \& Innanen, 1991; Caranicolas \& Papadopoulos, 2003). Finally, we introduce a new dynamical indicator which represents the orbital frequency of a star, which is defined as $f = \upsilon _{tot}/d$. Figure 12a-d shows the above four indicators for the regular orbit of Fig. 7a. One can observe, that all profiles are symmetrical with no sudden fluctuations, which is characteristic of a regular, quasi-periodic orbit. As one can see in Figure 12c the angular momentum of this orbit nicely oscillates symmetrically between two values of opposite sign and would therefore have a time average angular momentum equal to zero. Also, it is obvious from Figure 7a, that the orbit has clear turning points and this characterizes it as a box(let). On the contrary, things are very different in Figure 13a-d, which shows the same profiles but for the chaotic orbit of Fig. 8a. First the profiles of the distance (Fig. 13a) and total velocity (Fig. 13b) display a large number of asymmetries and abrupt changes. Furthermore, the time evolution of $L_z$ (Fig. 13c) is quite similar. Last but not least, the orbital frequency $f$ has a profile (Fig. 13d) with large and small peaks as time evolves. The main conclusion is that all four indicators coincide and prove that the orbit of Fig. 8a is chaotic.

It is interesting to note that there are two additional chaotic layers in Fig. 6. The inner chaotic layer surrounds the set of four islands in Fig. 6, while the outer layer surrounds all the set of the invariant curves. Thus in this case, we have a potential producing accurate periodic orbits and three different chaotic components as well (see Saito \& Ichimura 1979). This phenomenon was not observed in Caranicolas (2000), where we had a potential producing exact periodic orbits and a unified chaotic sea. As expected, the Lyapunov Characteristic Exponents (LCEs) in each chaotic region are different. In the central chaotic region (C.R) the LCE was found to be 0.127, in the inner chaotic layer (C.L-I) is 0.028, while in the outer chaotic layer (C.L-II) was found to be 0.083. In Figure 14 we can observe the time evolution of all LCEs of the dynamical system.
\begin{figure}[!tH]
\centering
\resizebox{0.60\hsize}{!}{\rotatebox{0}{\includegraphics*{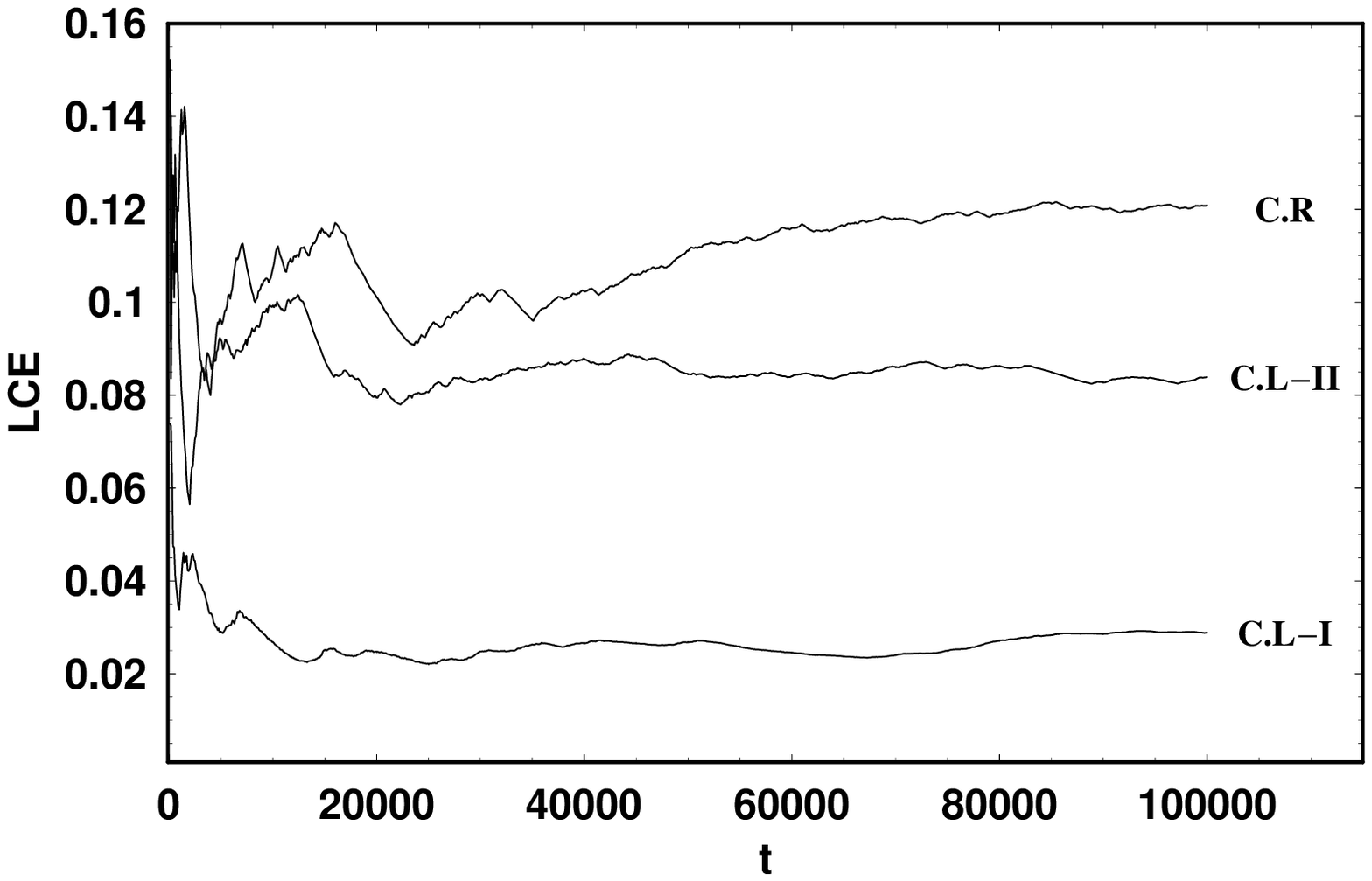}}}
\caption{Time evolution of the LCEs of the chaotic region (C.R) and the two chaotic layers (C.L-I, C.L-II).}
\end{figure}

Of course an extensive numerical investigation of the potential (2) for a large range of the constants $A$a and $B$ and also the parameter $c$ would be out of the scope of this paper. We believe that, we have presented two interesting cases, where the parameters of the potential (2) are given by equation (9) the total energy of the test-particle is given by equation (10) and (23) also is valid. Then the potential (2) possesses accurate elliptic periodic orbits and describes the central parts of a barred galaxy. As we see there are cases, where the phase plane is regular and cases, where the phase plane displays accurate periodic orbits and chaotic components as well.

In order to check the stability of the central invariant point, we carried out numerical experiments when $B=-1$ and $c=0.24$ for different values of $A$. Our numerical calculations indicate that the central periodic point is stable when $A\leq 2.2$. Note that always $A>1$. The stability of the central periodic point can be explained if we use the relation
\begin{equation}
\phi(y,p_y) = \left(3.02 - A\right)p_y^2 + \left(y + \frac{B}{2}\right)^2 - A y^2 p_y^2.
\end{equation}
Relation (27) was found, using a combination of mathematical analysis along with numerical simulations of the dynamical system and gives the structure of the $(y, p_y)$ phase plane near the central invariant point. It was found, that this relation gives a stable central periodic point when $A \leq 2.2, |B| \leq 1.0$. The corresponding value of the energy given by (10) is $E_{el}=0.740291$. Here we must note that the accuracy of the relation (27) is for the above given critical values of $A$, $B$ and $c$. On the other hand, the accuracy of the numerical calculations is better, almost $\pm 0.01$. This means that we go from $A=2.2$, where the central periodic point is stable to $A=2.3$, where it is unstable according to (27) and from $A=2.2$, where the central periodic point is stable to $A=2.21$, where it is unstable according to numerical integration. The same is true for $B$ and $c$. Generally speaking for each pair of the values of the two of the three parameters $A$, $B$ and $c$ there is a value of the third parameter that gives a value of energy (10). For this value of the energy and higher values the central invariant point is stable, while for lower values of the energy it is unstable. Figure 15a shows the topology of the $(y, p_y)$ phase plane near the central invariant point when $A=2$, $B=-1$ and $c=0.24$, while Figure 15b is similar to 15a when $A=4$, $B=-1$ and $c=0.24$. One can see, that relation (27) gives a representative picture of the topology of the phase plane near the central invariant point.
\begin{figure*}[!tH]
\centering
\resizebox{0.90\textwidth}{!}{\rotatebox{0}{\includegraphics*{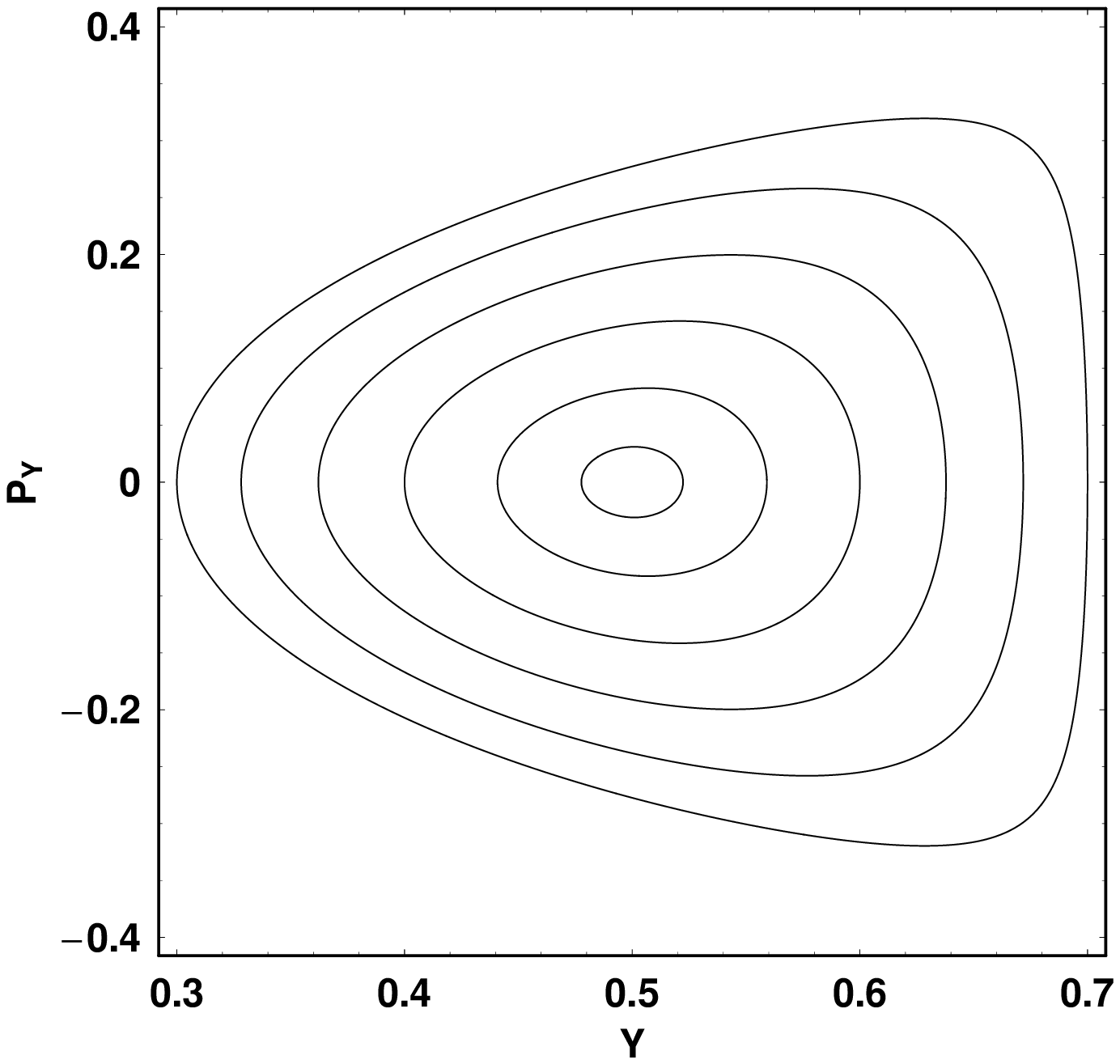}}\hspace{1cm}
                              \rotatebox{0}{\includegraphics*{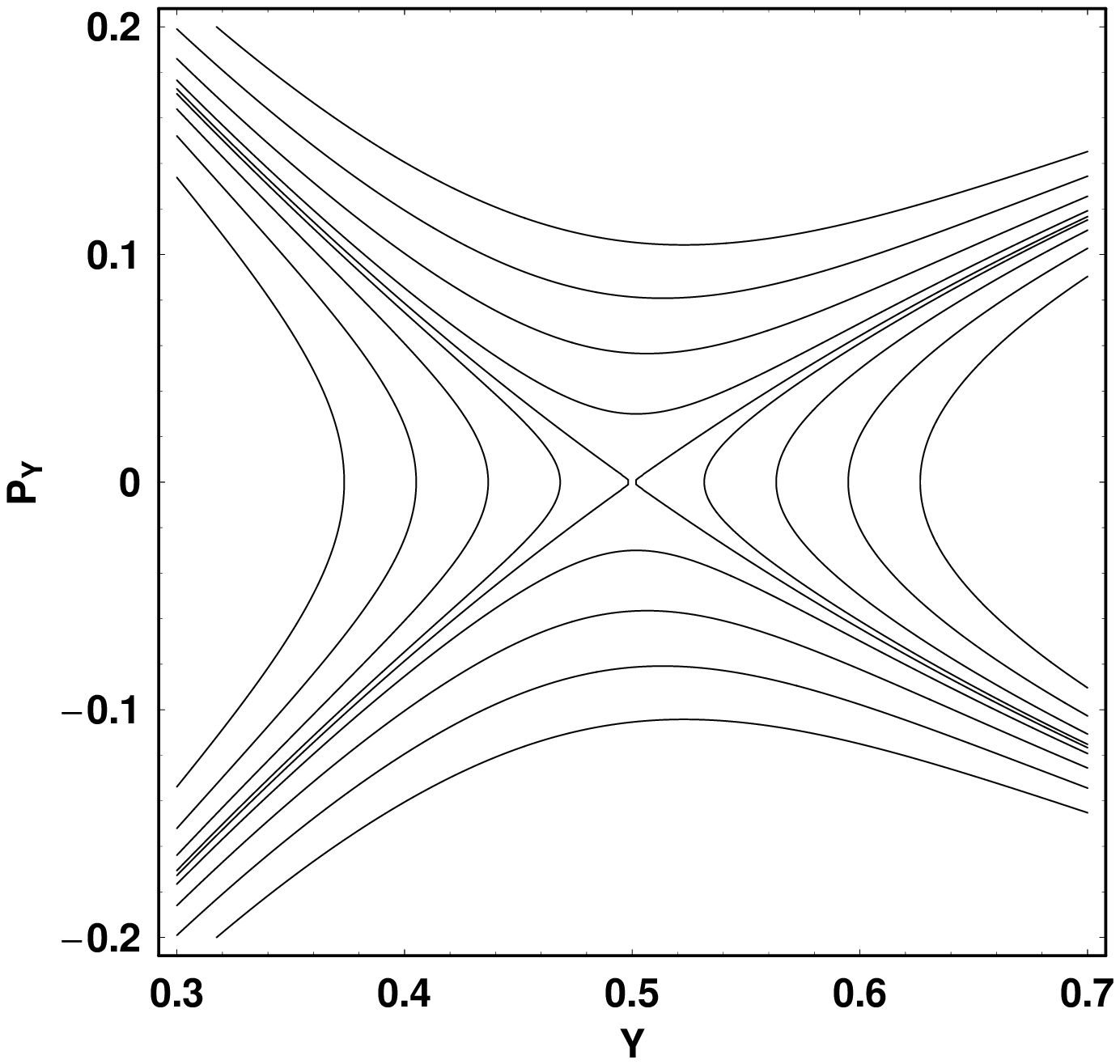}}}
\vskip 0.01cm
\caption{(a-b): The topology near the central invariant point. (a-left): $A=2$, $B=-1$, $c=0.24$ and (b-right): $A=4$, $B=-1$, $c=0.24$.}
\end{figure*}

One interesting question one might ask is the following: How sensitive is the the dynamical system when perturbing the obtained coefficients (9)? In order to answer this we have changed all the coefficients (9) equally in the range $0.01 \leq \Delta
\sigma_i/ \sigma_i \leq 0.1$, where $\sigma_i$ is any of $b, \alpha_1,...,\alpha_5$, while keeping the energy constant $E_{el}=0.3492$, when $A=4$ and $B=-1$. The results are not surprising. What happens is that $P1$ and $P2$ do not represent
accurate periodic orbits now. Furthermore, the chaotic region of Fig. 6 does not show any significant changes. Similar results
are found for the phase plane given in Fig. 3.
\begin{figure}[!tH]
\centering
\resizebox{0.60\hsize}{!}{\rotatebox{270}{\includegraphics*{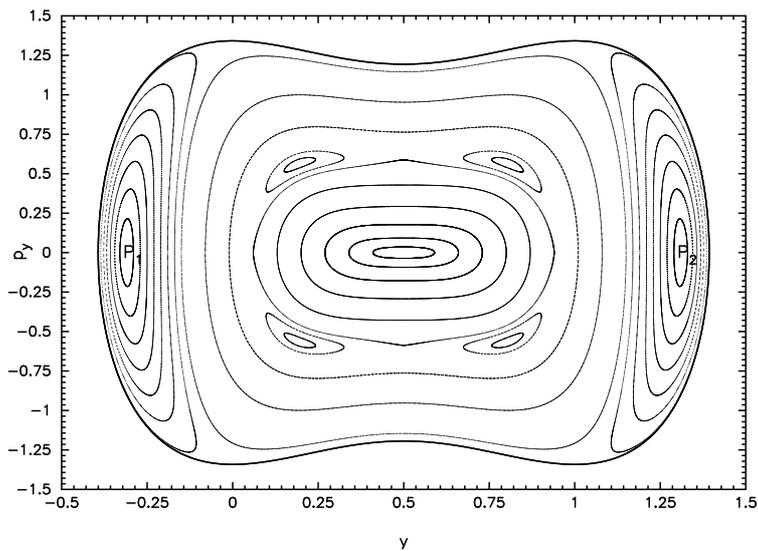}}}
\caption{The $(y, p_y)$ phase plane when $A=4$, $B=-1$ and $E=0.90$. $P1$ and $P2$ indicate the position of the non-accurate periodic orbits.}
\end{figure}

What is more interesting is to leave the coefficients (9) unchanged and increase the energy to values $E > E_{el}$. We have made the relevant calculations when $A=4$ and $B=-1$. The results are interesting. For values of the energy in the range $0.342 \leq E \leq 0.700$ the central periodic point is unstable, while the chaotic region near it decreases gradually, so that when $E=0.70$ it is negligible. For values of the energy in the range $0.700 < E \leq 0.871$ the central periodic point remains unstable. When $E \geq 0.872$ the central periodic point becomes stable. Furthermore, as the energy increases $P1$ and $P2$ still exist and move outwards but, of course, they do not represent accurate periodic orbits anymore. This behavior is quite different from that observed in Caranicolas (2000), where an increase of energy for the potential $V_P$ leads to larger chaotic regions. Figure 16 shows the $(y, p_y)$ phase plane when $A=4$, $B=-1$ and $E=0.90$. $P1$ and $P2$ indicate the position of the non-accurate periodic orbits.

\section{Discussion and conclusions}

In the present article, we have constructed a polynomial potential producing a mono-parametric family of ellipses following the theory of the Inverse Problem of Dynamics (IPD). The basic reason for considering a polynomial potential is that such a potential, generally, produces families of accurate periodic orbits, for some set of the values of the parameters involved for appropriate values of the energy. This does not happen, or it is not easy, to other types of galactic potentials, such as galactic mass models or galactic logarithmic models (see Binney \& Tremaine, 2008). It is also evident, that the outcomes of this work do not apply to a barred galaxy as a whole but only in the central region of a non-rotating isolated bar. Potentials describing motion in barred galaxies have been found on the context of the inverse problem in earlier papers (Caranicolas, 1998; Caranicolas \& Karanis, 1999). The above local potentials do not have accurate but only approximate periodic solutions. A first look at Fig. 1 gives the impression that we are dealing with the standard potential of a rotating bar, but all the structure we see in Fig. 1, is due to a combination of the gravitational forces with the effects that arise in a rotating reference system. This is because of the third and fourth order terms (especially $x^2y$ and $x^2y^2$) in equation (2).

The family of accurate periodic orbits produced by the polynomial potential is based on two constants $A$ and $B$ and one parameter $c$. Given the values of $A$ and $B$, $c$ must fulfill relation (23) in order our potential to represent motion in a barred galaxy. In this case, we have a potential for the central parts of a barred galaxy producing accurate stable elliptic periodic orbits. Near the elliptic periodic orbits there are families of trapped quasi periodic orbits supporting the barred structure. Another interesting family produced by our potential is the family of the box orbits, which also support the barred structure.

What is of particular interest is that there are cases, where our polynomial potential produces not only regular but also chaotic orbits. This happens when the central invariant point is unstable. Here, we must emphasize that the behavior of the
polynomial potential $V_G$ is quite different from that of the potential $V_P$ used by Caranicolas (2000), where we had a unified chaotic sea and not distinct chaotic components. In Section 4, we have used different types of dynamical indicators, in order to characterize the nature of motion. These indicators are extremely useful because they are very fast, due to the
fact, that they need time period of only 100 time units. On the contrary, the classical method of the LCE needs periods of order $10^5$ time units in order to obtain reliable and conclusive results. Moreover, the $P(f)$ indicator can also be used in order to obtain reliable results, regarding the character of an orbit. As for the $S(g)$ spectrum, the results lead to the conclusion, that it is a very useful dynamical indicator, in order to distinguish the regular or chaotic motion and also to identify resonant orbits, producing sets of islands of invariant curves in the surface of section. The $S(g)$ spectrum is faster than the $S(\alpha)$ spectrum, used by Caranicolas and Vozikis (1999), because it needs only one orbit and a small number of iterations $10^3 - 10^4$. Moreover, the $S(\alpha)$ spectrum, has vary limited ability to detect islandic and sticky motion. All these dynamical indicators can be easily extended in order to characterize the nature of orbits in 3D galactic potentials.

This and our previous papers, mentioned above, are part of our effort to connect the Inverse Problem to Galactic Dynamics. We believe that in the future we will be able to present new and more interesting results relating the Inverse Problem of Dynamics and realistic 3D galactic potentials. But the real galactic potentials describe the dynamical properties of galaxies and, therefore, they are connected to observational data. On this basis, it is evident that a connection of the astronomical observations with the Inverse Problem can be established. It would be of great interest to use data from observations, as a source, in order to construct 3D potentials describing real galaxies. Let us hope that we will manage to reach this target in the future.

\end{document}